\documentclass[aps,superscriptaddress,showpacs,nofootinbib,floatfix,epfs]{revtex4}
\usepackage{amsfonts,amscd,amsmath,amssymb,graphicx,color}
\usepackage[section]{placeins}
\def\xm{x_{\scriptscriptstyle -}}
\def\xp{x_{\scriptscriptstyle +}}
\def\rm{r_{\scriptscriptstyle -}}
\def\rp{r_{\scriptscriptstyle +}}
\def\am{\alpha_{\scriptscriptstyle -}}
\def\ap{\alpha_{\scriptscriptstyle +}}
\def\nup{\nu_{\scriptscriptstyle +}}
\def\num{\nu_{\scriptscriptstyle -}}
\def\nuz{\nu_{\scriptscriptstyle 0}}

\def\lam{\lambda_{\scriptscriptstyle -}}
\def\la0{\lambda_{\scriptscriptstyle 0}}
\def\nun{\nu_{\scriptscriptstyle 0}}
\def\lap{\lambda_{\scriptscriptstyle +}}
\def\qqb{\text{\tiny Q}\bar{\text{\tiny Q}}}
\def\qq{\text{\tiny QQ}}

\def\3q{3\text{\tiny Q}}
\def\qd{\text{\tiny QD}}
\def\kd{\kappa_{\text{\tiny d}}}
\def\oh{\frac{1}{2}}

\def\ep{\text{e}}
\def\g{\mathfrak{g}}
\def\oh{\frac{1}{2}}

\def\s{\mathfrak{s}}
\begin{document}
\title{Some Aspects of Three-Quark Potentials}
\author{Oleg Andreev}
 \affiliation{L.D. Landau Institute for Theoretical Physics, Kosygina 2, 119334 Moscow, Russia}
\affiliation{Arnold Sommerfeld Center for Theoretical Physics, LMU-M\"unchen, Theresienstrasse 37, 80333 M\"unchen, Germany}
%\date{}
\begin{abstract} 
We analytically evaluate the expectation value of a baryonic Wilson loop in a holographic model of an $SU(3)$ pure gauge theory. We then discuss three aspects of a static three-quark potential: an aspect of universality which concerns properties independent of a geometric configuration of quarks; a heavy diquark structure; and a relation between the three and two-quark potentials. 

\end{abstract}
\preprint{LMU-ASC ???}
\maketitle

\vspace{-7.5cm}
\begin{flushright}
LMU-ASC 68/15
\end{flushright}
\vspace{6cm}

%__________________________________________________________  1
\section{Introduction} 
\renewcommand{\theequation}{1.\arabic{equation}}
\setcounter{equation}{0}

Since many years there has been a great interest in triply heavy baryons \cite{bj,richard}. The challenge here is to reach the level of knowledge similar to that of charmonium and bottomonium. On the theoretical side, the challenge is to explain the structure and properties of such baryons. It is expected that the potential models would be useful in doing so. Furthermore, in the process one would expect to gain important insights into understanding how baryons are put together from quarks. 

The three-quark potential is one of the most important inputs of the potential models and also a key to understanding the quark confinement mechanism in baryons. However, so far, there is no reliable formula to describe the three-quark potential. The best known phenomenological models are ans\"atze which are a kind of the Cornell model for a three-quark system. The $\Delta$-law states that the static three-quark potential is a sum of two-quark potentials \cite{delta-law}.\footnote{For more details, see the Appendix D.} In practice, it is taken to be the sum of Cornell-type potentials

\begin{equation}\label{delta}
E_{\3q}(\mathbf{x}_1,\mathbf{x}_2,\mathbf{x}_3)=
\sum_{i<j}^3 
-\frac{\alpha_{\3q}}{\vert\mathbf{x}_{ij}\vert}+\oh\sigma \vert\mathbf{x}_{ij}\vert+C
\,,
\end{equation}
where $\mathbf{x}_i$ is the position vector of the $i$-th quark, $\mathbf{x}_{ij}=\mathbf{x}_i-\mathbf{x}_j$, $\alpha_{\3q}$ and $C$ are parameters, and $\sigma$ is the string tension. The relation with the quark-antiquark potential implies that $\alpha_{\3q}=\oh\alpha_{\qqb}$. The $\Delta$-law predicts that the energy grows linearly with the perimeter of the triangle formed by quarks. Alternatively, the $Y$-law states that the energy grows linearly with the minimal length of a string network which has a junction at the Fermat point of the triangle \cite{artu,isgur,cornwall}. In this case, a slight modification of \eqref{delta} is of common use in the potential models. It is simply

\begin{equation}\label{Y}
E_{\3q}(\mathbf{x}_1,\mathbf{x}_2,\mathbf{x}_3)= \sum_{i<j}^3 
-\frac{\alpha_{\3q}}{\vert\mathbf{x}_{ij}\vert}+\sigma L_{\text{\tiny min}}+C
\,.
\end{equation}

Unfortunately, there have not yet been any experimental results concerning triply heavy baryons. Thus, the predictions\footnote{For a recent development, see, e.g., \cite{pheno}.} based on either the $\Delta$-law or the $Y$-law can not be compared to the real world, and the last word has not yet been said on this matter. In such a situation lattice gauge theory is the premier method for obtaining quantitative and nonperturbative predictions from strongly interacting gauge theories. The three-quark potentials are being studied on the lattice \cite{review}. Although the accuracy of numerical simulations has been improved, the situation is still not completely clear. There is no problem with the fit by the $Y$-law at long distances \cite{suganuma3q,jahn3q,suganuma3q-new}. At shorter distances the conclusion is mixed in the sense that some results favor the $\Delta$-law \cite{pdf3q,jahn3q}, and others favor the $Y$-law \cite{suganuma3q,suganuma3q-new,koma}. 

The situation is even worse with the hybrid three-quark potentials. So far, there are no phenomenological predictions and very little is known about those from lattice studies.\footnote{See, however, \cite{suganuma-hybrid}.} These potentials remain almost unknown and, therefore, merit more attention, as in the case of the hybrid quark-antiquark potentials \cite{review,hybrid-review}.

One of the implications of the AdS/CFT correspondence \cite{malda} is that it opened a new window 
for studying strongly coupled gauge theories and, as a result, resumed interest in finding a string description (string dual) of QCD. It is worth noting that it is not an old fashioned four-dimensional string theory in Minkowski space, but a five (ten)-dimensional one in a curved space.

In this paper we continue a series of studies \cite{a-bar,a-Nq} devoted to the three-quark potentials within a five (ten)-dimensional effective string theory. The model we are developing represents a kind of soft wall model of \cite{son}, where the violation of conformal symmetry is manifest in the background metric \cite{q2}. It would be unwise to pretend that such a model is dual to QCD or that it can be deduced starting from the AdS/CFT correspondence. Our reasons for now pursuing this model are:

(1) Because there still is no string theory which is dual to QCD. It would seem very good to gain what experience we can by solving any problems that can be solved with the effective string model already at our disposal. The AdS/CFT correspondence is a good starting point, because we already know a lot about Wilson loops in ${\cal N}=4$ supersymmetric Yang-Mills theory \cite{ads}.\footnote{There is an obvious question. If $N=3$, why is it a good starting point? We have no answer and take this as an assumption, but (2) partially resolves the question.}

(2) Because the results provided by this model are consistent with the lattice calculations and QCD phenomenology. In some cases the quantitative agreement is so good that one could not even expect such a success from a simple model. In \cite{az1}, we have computed the quark-antiquark potential. Subsequent work  made it clear that the model should be taken seriously, particular in the context of consistency with the lattice \cite{white} and quarkonium spectrum \cite{carlucci}. Another instance of perfect consistency between the model \cite{a-Nq} and the lattice calculations \cite{pdf3q,suganuma3q,suganuma3q-new} is the three-quark potential obtained on an equilateral triangle. In addition, the model also reproduces one of the hybrid potentials of \cite{kuti}. It is the $\Sigma_u^-$ potential \cite{hybrids}.

(3) Because analytic formulas are obtained by solving the model. This allows one to easily compare the results with the lattice and QCD phenomenology. 

(4) Because the model is able to explore QCD properties in the transition region between confinement and asymptotic freedom. As seen from Figure 3, it can operate at length scales down to $0.2\,\text{fm}$. This is a big advantage of the model over any type of an old fashioned four-dimensional string model in Minkowski space. 

(5) Because the aim of our work is to make predictions which may then be tested by means of other methods, e.g., numerical simulations.

Of course, it is worth keeping in mind that this model, as any other model, has its own limitations. In particular, it breaks down at very small length scales that makes impossible to compare the results with those of perturbative QCD.

The paper is organized as follows. For orientation, we begin by setting the framework and recalling 
some preliminary results. Then, we consider the three-quark potential of collinear quarks. This allows us to compare the results with those obtained on the equilateral geometry and make 
predictions on what is expected to be universal (independent of a geometrical configuration of quarks) 
at short and long distances. We go on in Section III to discuss the quark-diquark and quark-quark potentials as the limiting cases of that obtained for the collinear geometry. Our goal here is to determine the leading terms in the interquark potential. In Section IV, we give an example of the three-quark hybrid potential. Here we also consider a relation between two and three hybrid quark potentials and 
make a prediction of universality for a gap between the potential and its hybrid at long distances. We conclude in Section V with a discussion of some open problems and possibilities for further study. Some technical details are given in the Appendices.

%__________________________________________________________________________________________
\section{Three-Quark Potential via Gauge/String Duality}
\renewcommand{\theequation}{2.\arabic{equation}}
\setcounter{equation}{0}

In this section we will derive the three-quark potential for two different geometries. We start with 
an equilateral triangle geometry and then consider a collinear geometry. Although we mainly concentrate on the collinear geometry, as our basic example, the approach is equally applicable for any geometry. The Appendices A and B provide the necessary toolkit for doing so.

%________________________________________________________
\subsection{Preliminaries}

\begin{figure}[htbp]
\centering
\includegraphics[width=4.25cm]{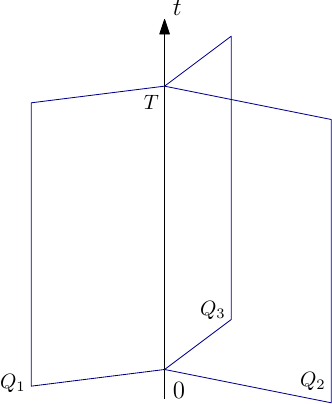}
\caption{{\small A baryonic Wilson loop. A three-quark state $Q_1Q_2Q_3$ is generated at $t=0$ and is annihilated at $t=T$.}}
\label{wfig}
\end{figure}

The static three-quark potential can be determined from the expectation value of a baryonic Wilson loop \cite{review}. The Wilson loop in question is defined in a gauge-invariant manner as $W_{\3q}=\tfrac{1}{3!}\varepsilon_{abc}\varepsilon_{a'b'c'}U_1^{aa'}U_2^{bb'}U_3^{cc'}$, with $U_i$ the path-ordered exponents along the lines shown in Figure \ref{wfig}. 
%________________________  f - 1 __________________________________
%______________________________________________________________________
In the limit $T\rightarrow\infty$ the expectation value of $W_{\3q}$ is given by 

\begin{equation}\label{bloop}
\langle W_{\3q}\rangle=\sum_{n=0}^\infty B_n\ep^{-E_{\3q}^{(n)}T}
\,.
\end{equation}
Here $E_{\3q}^{(0)}$ is called the three-quark potential (ground state energy) if the corresponding contribution dominates the sum as $T$ approaches infinity. In other words, it requires $E_{\3q}^{(0)}<E_{\3q}^{(i)}$ for any $i>0$ and any quark configuration. If so, then the remaining $E_{\3q}^{(i)}$'s are called hybrid three-quark potentials (excited state energies).

In our study of baryonic Wilson loops, we adapt a formalism proposed within the AdS/CFT correspondence \cite{witten,gross} to a confining gauge theory. 

First, we take the ansatz for the background metric

\begin{equation}\label{metric10}
ds^2=\ep^{\s r^2}\frac{R^2}{r^2}\bigl(dt^2+d\vec x^2+dr^2\bigr)+\text{e}^{-\s r^2}g_{ab}^{(5)}d\omega^a d\omega^b 
\,.
\end{equation}
Thus, the geometry in question is a one-parameter deformation, parameterized by $\s$, of a product of the Euclidean $\text{AdS}_5$ space of radius $R$ and a 5-dimensional compact space (sphere) $\mathbf{X}$ whose coordinates are $\omega^a$. In \eqref{metric10}, there are the two free parameters $R$ and $\s$. Note that the first combines with $\alpha'$ from the Nambu-Goto action \eqref{NG} such that $\g=\frac{R^2}{2\pi\alpha'}$ then is to be fitted. Since we deal with an effective string theory based on the Nambu-Goto 
formulation \footnote{This has some limitations. The reason is that the Green-Schwarz formulation similar to that for $\text{AdS}_5\times\mathbf{S}^5$ is still missing. So, we consider the model as an effective theory rather than an ultimate solution for QCD.}, knowing the background geometry is sufficient for our purposes. There are good motivations for taking this ansatz. First, such a deformation of $\text{AdS}_5$ leads to linear Regge-like spectra for mesons \cite{son,q2} and 
the quark-antiquark potential \cite{az1} which is very satisfactory in the light of lattice gauge theory and phenomenology \cite{white,carlucci}. Second, the deformation of $\mathbf{X}$ is motivated by thermodynamics \cite{a-pis}.

Next we consider the baryon vertex. In static gauge we take the following ansatz for its action

\begin{equation}\label{baryon-v}
S_{\text{vert}}=m\frac{\ep^{-2\s r^2}}{r}T\,,
\end{equation}
where $m$ and $\s$ are parameters. Note that the parameter $\s$ is the same as in \eqref{metric10} so that only $m$ is a new one.

In what follows we will assume that quarks are placed at the same point in the internal space.\footnote{It is worth noting that this assumption makes the problem effectively five-dimensional and hence more tractable. From the five-dimensional point of 
view, the vertex looks like a point-like object (particle).} Therefore the detailed structure of $\mathbf{X}$ is not important, except the exponential warp factor depending on the radial direction. The motivation for such a form of the warp factor is drawn from the AdS/CFT construction, where the baryon vertex is a five-brane \cite{witten}. Taking a term $\int dtd^5\omega\sqrt{g^{(6)}}$ from the world-volume action of the brane results in $T\ep^{-2\s r^2}/r$. This is, of course, a heuristic  argument but it leads to the very satisfactory result that the lattice data of \cite{pdf3q,suganuma3q,suganuma3q-new} obtained on an equilateral triangle can be described by a single parameter \cite{a-Nq}.

One of the important differences between QCD and AdS/CFT is that QCD has got much further than AdS/CFT in addressing the issue of hybrid potentials \cite{hybrid-review}. The common wisdom is that those potentials correspond to excited 
strings \cite{isgur}. The structure of string excitations is quite complicated and is made of many different kinds of elementary excitations like vibrational modes, loops, knots etc. Adopting the view point that some excitations imply a formation of cusps, one can 
model them by inserting local objects (defects) on a string.\footnote{A similar idea was used in four-dimensional string models, but with a different goal, such as a description of linear baryons. For more discussion and references, see \cite{nesterenko}.} Thus, what we need is an action for such an object. In static gauge we take the action to be of the form

\begin{equation}\label{defect-d}
S_{\text{def}}=n\frac{\ep^{-2\s r^2}}{r}T
\,,
\end{equation}
where $n$ and $\s$ are parameters. Again, $\s$ is the same as in \eqref{metric10}. 

The form like \eqref{defect-d} seems natural if one thinks of the defect as a tiny loop formed by a pair of baryon-antibaryon vertices connected with fundamental strings. In such a scenario, the expression \eqref{defect-d} is based on an assumption that the strings do not change the radial dependence in \eqref{baryon-v}. This is also a heuristic argument but it can be confronted with the lattice data of \cite{kuti}. For $\Sigma_u^-$, the result is in good agreement \cite{hybrids}. Notice that \eqref{defect-d} contains only one free parameter that makes it attractive from the phenomenological point of view.

Finally, we place heavy quarks at the boundary points of the five-dimensional space ($r=0)$ but at the same point in the internal space $\mathbf{X}$. We consider configurations in which each quark is the endpoint of the Nambu-Goto string, with the strings join at the baryon vertex in the interior as shown in Figure \ref{etfig}. For excited strings, we also place defects on them as shown in Figure \ref{hybfig}. The total action of a system has, in addition to the standard Nambu-Goto actions, also contributions arising from the baryon vertex and defect. The expectation value of the Wilson loop is then

\begin{equation}\label{W3}
\langle W_{\3q}\rangle\sim\ep^{-S_{\text{\tiny min}}}
\,,
\end{equation}
where $S_{\text{\tiny min}}$ is the minimal action of the system. Combining it with $\eqref{bloop}$ gives the the three-quark potential.\footnote{Like in AdS/CFT, this formula is oversimplified for various reasons, but it seems acceptable for the purposes of the effective string theory based on the Nambu-Goto formulation.}

%__________________________________________________________
\subsection{Equilateral triangle geometry}

\begin{figure}[htbp]
\centering
\includegraphics[width=5.5cm]{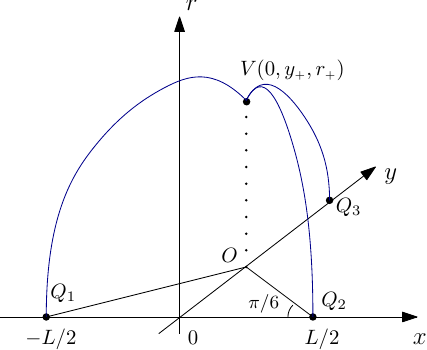}
\caption{{\small A baryon configuration. The quarks $Q_i$ are placed at the vertices of the equilateral triangle. $V$ is a baryon vertex.}}
\label{etfig}
\end{figure}

As a warmup, let us give a derivation of the three-quark potential in the case when the quarks are at the vertices of an equilateral triangle of length $L$ \cite{a-Nq}. To this end, we consider the configuration shown in Figure \ref{etfig}. Here, gravity pulls the baryon vertex toward the boundary that allows the result to be consistent with the lattice data \cite{pdf3q,suganuma3q,jahn3q,suganuma3q-new}.

The $D_3$ symmetry of the problem immediately implies that the projection of $V$ onto the $xy$ plane is a center of the triangle and all the strings have an identical profile. In this case, a radius of the circumscribed circle is given by \eqref{l-} and, as a consequence, the triangle's side length is\footnote{We abbreviate $\nup$ to $\nu$ when this is not ambiguous.}

\begin{equation}\label{etL}
L=\sqrt{\frac{3\lambda}{\s}}
\biggl[
\int^1_0 dv\, v^2\, \ep^{\lambda(1-v^2)}\Bigl(1-v^4\ep^{2\lambda(1-v^2)}\Bigr)^{-\frac{1}{2}}+
\int^1_{\sqrt{\frac{\nu}{\lambda}}} dv\, v^2\, \ep^{\lambda(1-v^2)}\Bigl(1-v^4 \ep^{2\lambda(1-v^2)}
\Bigr)^{-\frac{1}{2}}
\biggr]\,,
\end{equation}
where $\lambda\in[0,1]$ and $\nu\leq\lambda$. Note that $\nu=\s\rp^2$ and $\lambda=\s r_{\text{\tiny max}}^2$, with $r_{\text{\tiny max}}$ shown in Figure \ref{afigs}.

%________________________  f - 2 __________________________________
%________________________

It is straightforward to compute the total energy of the configuration $E=\sum_{i=1}^3 E_i +{\cal V}(\rp)$. The energy of a single string is given by \eqref{E-}. Then, using this expression and the 
expression \eqref{baryon-v} for the gravitational energy of the baryon vertex, we find 

\begin{equation}\label{etE}
E_{\3q}=3\g\sqrt{\frac{\s}{\lambda}}
\biggl[
\kappa\sqrt{\frac{\lambda}{\nu}}\ep^{-2\nu}+
\int^1_0\,\frac{dv}{v^2}\,\biggl(\ep^{\lambda v^2}\Bigl(1-v^4\ep^{2\lambda (1-v^2)}\Bigr)^{-\frac{1}{2}}
-1-v^2\biggr)+
\int^1_{\sqrt{\frac{\nu}{\lambda}}}\,\frac{dv}{v^2}\,\ep^{\lambda v^2}\Bigl(1-v^4\,\ep^{2\lambda (1-v^2)}\Bigr)^{-\frac{1}{2}}
\biggr]+C
\,,
\end{equation}
with $C$ a normalization constant. It is equal to $3c$ as it follows from \eqref{E-}. In addition, $\kappa=\frac{m}{3\g}$, as defined in \eqref{et-netforce}.

At this point, we can complete the parametric description of the three-quark potential for the equilateral triangle geometry. To this end, the gluing conditions at the vertex should be involved. When the triangle is equilateral, we impose the condition \eqref{et-netforce} 

\begin{equation}\label{et-noforce}
\sin\alpha=\kappa(1+4\nu)\ep^{-3\nu}
\end{equation}
which is nothing else but the balance of force in the radial direction. Here we have abbreviated $\ap$ to $\alpha$. Note that a negative value of $\alpha$, as sketched in Figure \ref{etfig}, implies that $\kappa<0$. In this case, the gravitational force is directed in the downward vertical direction.

Combining \eqref{et-noforce} with \eqref{lambda}, one can express $\lambda$ in terms of the ProductLog 
function \cite{wolf}

\begin{equation}\label{etlambda}
\lambda(\nu)=-\text{ProductLog}[-\nu\ep^{-\nu}(1-\kappa^2(1+4\nu)^2\ep^{-6\nu})^{-\frac{1}{2}}]
\,.
\end{equation}
Here $\nu\in[0,\nu_{\ast}]$, with $\nu_\ast$ a solution to $\lambda(\nu)=1$.

Thus the potential is given in parametric form by $E_{\3q}=E(\nu)$ and $L=L(\nu)$. The parameter takes values on the interval $[0,\nu_{\ast}]$. This is the main result of this section. It is a special case of the result announced in \cite{a-Nq}.

Now we wish to compare this result to the lattice data and see some of the surprising features of our model. But first let us look at the behavior of $E_{\3q}(L)$ at short and long distances.

A simple analysis shows that $L(\nu)$ is a monotonically increasing function on the interval $[0,\nu_\ast]$, and that $L(\nu)$ goes to zero as $\nu\rightarrow 0$ and goes to infinity as $\nu\rightarrow\nu_\ast$. Thus, to analyze the short distance behavior of $E_{\3q}$, we need to find the asymptotic behavior of $L(\nu)$ and $E(\nu)$ near $\nu=0$.\footnote{For details, see the Appendix C.} In this case, we restrict ourselves to the two leading terms that allows us to easily obtain the energy as a function of the triangle's side length. Thus, at short distances the three-quark potential is given by 

\begin{equation}\label{etEL-small}
E_{\3q}(L)=-3\frac{\alpha_{\3q}}{L}+C+\frac{3}{2}\sigma_0 L+o(L)\,,
\end{equation}
with 
\begin{equation}\label{etalpha-sigma}
\alpha_{\3q}=-\frac{1}{3}L_0E_0\g
%-\frac{\sqrt{3}}{4}{\cal B}\bigl(\kappa^2;\tfrac{1}{2},\tfrac{3}{4}\bigr)
%\Bigl(\kappa\rho^{-\tfrac{1}{4}}+\frac{1}{4}
%{\cal B}\bigl(\kappa^2;\tfrac{1}{2},-\tfrac{1}{4}\bigr)\Bigr)\g
\,,\qquad
\sigma_0=\frac{2}{3L_0}\Bigl(E_1+\frac{L_1}{L_0}E_0\Bigr)\g\,\s
\,.
\end{equation}
The $L_i$'s and $E_i$'s are expressed in terms of the beta function and given by equations \eqref{l0e1} and \eqref{l0e2}, respectively.

In a similar spirit, we can explore the long distance behavior of $E_{\3q}$. Expanding the right hand sides of 
Eqs.\eqref{etL}-\eqref{etE} near $\nu=\nu_\ast$, we reduce these equations to a single equivalent equation  

\begin{equation}\label{etEL-large}
E_{\3q}(L)=\sqrt{3}\sigma L+{\text c}+o(1)\,,
\end{equation}
with 
\begin{equation}\label{tsigma-c}
\sigma=\ep\g\s
\,,\quad
{\text c}=3\g\sqrt{\s}\biggl[
\frac{\kappa}{\sqrt{\nu_\ast}}\ep^{-2\nu_\ast}+
\int_0^1\frac{dv}{v^2}\biggl(\ep^{v^2}\Bigl(1-v^4\ep^{2(1-v^2)}\Bigr)^{\frac{1}{2}}-1-v^2\biggr)
+
\int_{\sqrt{\nu_\ast}}^1\frac{dv}{v^2}\ep^{v^2}\Bigl(1-v^4\ep^{2(1-v^2)}\Bigr)^{\frac{1}{2}}
\biggr]
+C\,.
\end{equation}
Here $\sigma$ is the physical string tension. It remains universal and unaltered in all the cases: the quark-antiquark \cite{az1}, hybrid \cite{hybrids}, three-quark potentials \cite{a-bar}, and also in the examples we consider below. It is notable that the constant terms at short and long distances are different. Because of scheme ambiguities, each of those has no physical meaning, but the difference 

\begin{equation}\label{cC}
\text{c}-C=3\g\sqrt{\s}\biggl[
\frac{\kappa}{\sqrt{\nu_\ast}}\ep^{-2\nu_\ast}+
\int_0^1\frac{dv}{v^2}\biggl(\ep^{v^2}\Bigl(1-v^4\ep^{2(1-v^2)}\Bigr)^{\frac{1}{2}}-1-v^2\biggr)
+
\int_{\sqrt{\nu_\ast}}^1\frac{dv}{v^2}\ep^{v^2}\Bigl(1-v^4\ep^{2(1-v^2)}\Bigr)^{\frac{1}{2}}
\biggr]
\,,
\end{equation}
is not ambiguous and is free from divergences. This makes the model so different from the phenomenological laws \eqref{delta} and \eqref{Y}, where the difference vanishes.

\begin{figure}[htbp]
\centering
\includegraphics[width=8.3cm]{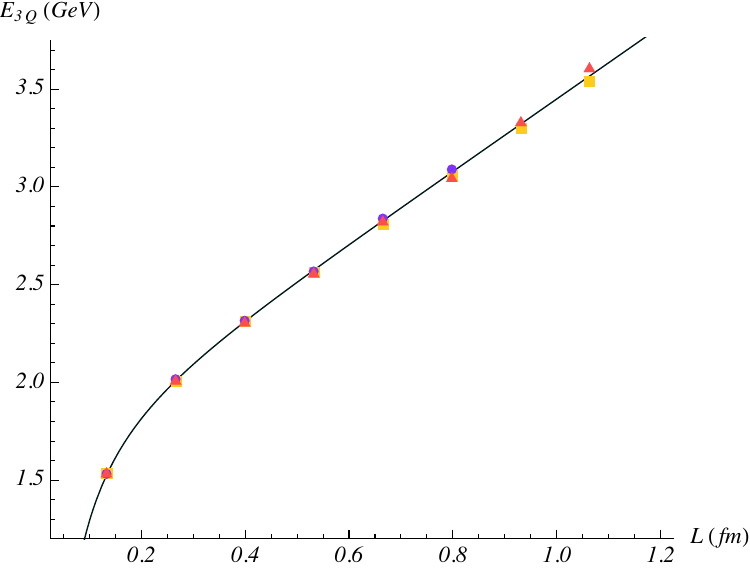}
\hspace{0.75cm}
\includegraphics[width=8.3cm]{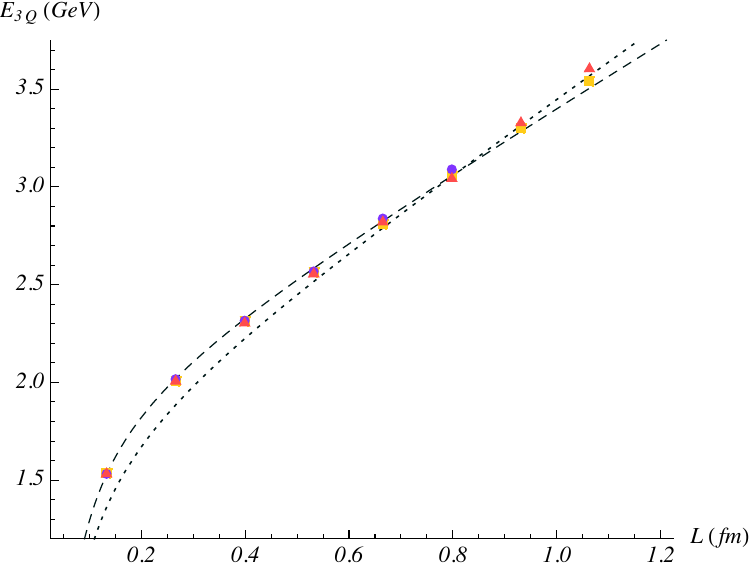}
\caption{{\small The lattice data are taken from \cite{pdf3q,jahn3q} (squares), \cite{suganuma3q} (disks), and \cite{suganuma3q-new} (triangles). We use the normalization of \cite{pdf3q}. We don't display any error bars because they are comparable to the size of the symbols. Left: $E_{\3q}$ as a function of $L$ at $\g=0.176$, $\s=0.44\,\text{GeV}^2$, $\kappa=-0.083$, and $C=1.87\,\text{GeV}$. Right: The $\Delta$-law \eqref{etEL-small} (dashed) and the Y-law \eqref{etEL-Y} (dotted).}}
\label{et-latticefig}
\end{figure}

Having found the asymptotic behaviors at short and long distances, we can compare our model of the three-quark potential with the results of numerical simulations. We proceed along the lines of \cite{a-Nq}. First, we set $\g = 0.176$ and $\s = 0.44\,\text{GeV}^2$, i.e., to the same values as those of \cite{hybrids} used for modeling the quark-antiquark potentials of \cite{kuti}. Then the remaining parameter is fitted to be $\kappa=-0.083$ using the data of \cite{pdf3q} from numerical simulations of the baryonic Wilson loops. The result is plotted in Figure \ref{et-latticefig}. 
%________________________  f - 2 __________________________________
%______________________________________________________________________
We see that the model reproduces the lattice data remarkably well with just one free parameter.

A simple estimate using \eqref{etalpha-sigma} and the fitted value of $\kappa$ shows that \cite{a-Nq}

\begin{equation}\label{etest}
\frac{\alpha_{\3q}}{\alpha_{\qqb}}\approx 0.495 
\,,\qquad
\frac{\sigma_0}{\sigma}\approx 1.007
\,,
\end{equation}
where $\alpha_{\qqb}$ is a coefficient in front of the Coulomb term of the quark-antiquark potential \cite{az1}. Explicitly, it is given by $\alpha_{\qqb}=(2\pi)^3\Gamma^{-4}\bigl(\tfrac{1}{4}\bigr)\g$ \cite{maldaq}. This suggests that at short distances \eqref{etEL-small} can be rewritten as 

\begin{equation}\label{etEL-small2}
E_{\3q}(L)\approx -\frac{3}{2}
\frac{\alpha_{\qqb}}{L}+C+\frac{3}{2}\sigma L+o(L)
\approx\frac{1}{2}\sum_{i<j}^3E_{\qqb}(r_{ij})
\,,
\end{equation}
where $E_{\qqb}(r_{ij})$ is the quark-antiquark potential and $r_{ij}$ denotes the distance between the vertices $i$ and $j$. With our choice of normalization, the normalization constant $C$ for the energy of a single baryon configuration is equal to $3c$, while that of the quark-antiquark pair  is $2c$ (see \eqref{E2}). This is consistent with the relation \eqref{etEL-small2}. Actually, it is the $\Delta$-law suggested in \cite{delta-law}. The analysis of \cite{pdf3q,jahn3q,a-Nq} shows that it is a good approximation to the lattice at distances shorter than $0.6-0.8\,\text{fm}$. 

The underlying physical picture is not necessarily very accurate for baryons. We suggest that the $\Delta$-law can be treated in a way that helps clarify the physics of strong interactions and at the same time complies with the lattice. The picture that emerges from this point of view is that at short distances the three-quark potential is described by a sum of two-body potentials. At length scales where a diquark can be treated as a point-like object, it is simply

\begin{subequations}
\begin{align}
E_{\3q}(L)\approx &\frac{1}{2}\sum_{i<j}^3E_{\qd}(r_{ij}) \label{etEL-small3}
\,, \\
\intertext{with $E_{\qd}$ a quark-diquark potential. This is plausible because in this case $E_{\qd}$ coincides with the quark-antiquark potential such that \eqref{etEL-small3} reduces to \eqref{etEL-small2}. Alternatively, it may be written as a sum over all quark-quark pairs}
E_{\3q}(L)\approx & \sum_{i<j}^3E_{\qq}(r_{ij}) \label{etEL-small3a}
\,,
\end{align}
\end{subequations}
with $E_{\qq}$ a quark-quark potential. Notice that \eqref{etEL-small3a} can be derived from \eqref{etEL-small2} by using the $\oh$-rule $E_{\qq}\approx\oh E_{\qqb}$ \cite{richard}. We will return to these issues again in Section III and the Appendix D. 

Actually, in connection with the above analysis, some additional questions should be asked. 
To reach the conclusion that in the interval $[0.2\,\text{fm},0.7\,\text{fm}]$ the three-quark potential is well approximated by a sum of two-body potentials we analyzed the short distance behavior of the expressions \eqref{etL} and \eqref{etE}. It turns out that the first terms of the series \eqref{etEL-small} provides a good approximation in this interval, as can be seen from Figure 3. This is a model artifact. Does it imply that \eqref{etEL-small} is also reliable at shorter distances? This is unclear. Certainly, no resummation of perturbative series of a pure $SU(3)$ gauge theory is known. But it might occur that resummation will lead to the form of \eqref{etEL-small}. In addition, the relation \eqref{etest} between $\alpha_{\qqb}$ and $\alpha_{\3q}$ is similar to that of the tree level result. What may be the reason? In the model we are considering $\alpha_{\3q}$ is a function of the 
parameter $\kappa$, while $\alpha_{\qqb}$ is a constant. So, there may be any relation between those coefficients. The peculiar one is the result of fitting to the lattice. Again, to really explain why it 
is so, resummation of perturbative series is needed.

What happens at longer distances? We can reasonably expect that \eqref{etEL-large} is a proper approximation for $E_{\3q}$. This is indeed the case \cite{a-Nq}. It is interesting that the L\"uscher correction to \eqref{etEL-large} turns out to be negligible. In fact, such a behavior is also provided by the Y-law at long distances \cite{artu,isgur}. 

Thus, the physical picture we have is that our model incorporates two-body interactions, the $\Delta$-law, at short distances and a genuine three-body interaction, the $Y$-law, at longer distances. In string context, the two-body interaction is described by a single string stretched between a quark and a diquark, while the three-body interaction is the standard one. It includes three strings meeting at a common junction. Mathematically, $E_{\3q}(L)$ is a complicated function of $L$ whose asymptotic behavior is described by the $\Delta$ and $Y$-laws.

If one includes the L\"uscher correction, like in the Y-law,\footnote{This issue was raised by H. Suganuma. Here we replace $\text{c}\approx 1.60\,\text{GeV}$ by $c_{\text{\tiny Y}}\approx 1.67\,\text{GeV}$ to better fit the data.}

\begin{equation}\label{etEL-Y}
E_{\3q}(L)=\sqrt{3}\sigma L+c_{\text{\tiny Y}}-3\frac{\alpha_{\3q}}{L}
\,,
\end{equation}
then the whole picture does not change and remains the same as before, except the transition between the two behaviors occurs at a slightly larger scale, of order $0.8\,\text{fm}$ as seen from Figure \ref{et-latticefig}.

So far, we have tacitly assumed that gravity pulls the vertex toward the boundary. This is 
an important difference between our model and those in the literature devoted to duals of large $N$ (supersymmetric) gauge theories. There are obvious questions one can ask about what took place. 
Is the $1/N$ expansion really a good approximation? What is the origin for $\kappa$ being negative? Is it a model artifact? Unfortunately, no real resolution of this problem will be proposed here. Our criterion is to mimic QCD and provide the basis for further calculations.
%__________________________________________________________ Symmetric Linear

\subsection{Symmetric collinear geometry}

Given the set of parameters that we have just fitted, it is straightforward to determine the three-quark potential for other geometries and make some predictions in the cases when there are no lattice data available. To get some intuition, we first consider the symmetric collinear geometry.

\begin{figure}[htbp]
\centering
\includegraphics[width=5.35cm]{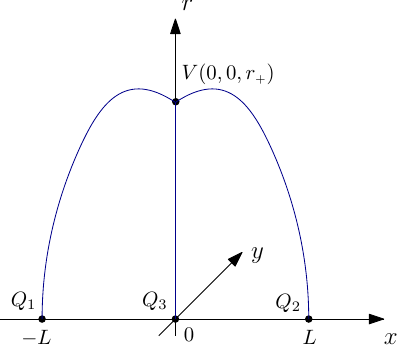}
\hspace{2.5cm}
\includegraphics[width=5.75cm]{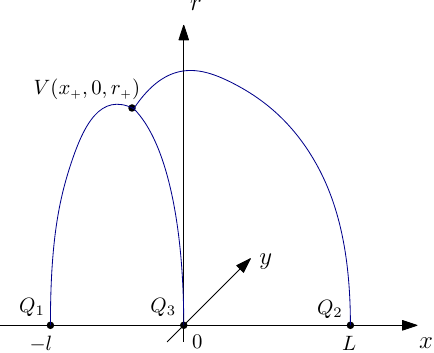}
\caption{\small{Typical collinear configurations. Left: A symmetric configuration. Right: A generic configuration. Here $l<L$.}}
\label{lfig}
\end{figure}

As before, we place the quarks at the boundary points of the five-dimensional space and consider a configuration in which each of the quarks is the endpoint of a fundamental string. The strings join at a baryon vertex in the interior as shown in Figure \ref{lfig}, on the left. 
%________________________  f - 2 __________________________________
%__________________________________________________________________
For convenience, the quarks are on the $x$-axis such that $x_1=-L$, $x_2=L$, and $x_3=0$.

Since the quark configuration is symmetric under reflection through $x = 0$, the side strings have an identical profile and the middle string is stretched in the radial direction. Given this, we can use the general formula \eqref{l-} to write

\begin{equation}\label{L-s}
L=\sqrt{\frac{\lambda}{\s}}\biggl[
\int^1_0 dv\, v^2\, \ep^{\lambda(1-v^2)}
\Bigl(1-v^4\,\ep^{2\lambda(1-v^2)}\Bigr)^{-\frac{1}{2}}+
\int^1_{\sqrt{\frac{\nu}{\lambda}}} 
dv\, v^2\, \ep^{\lambda(1-v^2)}
\Bigl(1-v^4\,\ep^{2\lambda(1-v^2)}\Bigr)^{-\frac{1}{2}}
\biggr]
\,,
\end{equation}
where $\nu=\s\rp^2$ and $\lambda=\s r_{\text{\tiny max}}^2$, with $r_{\text{\tiny max}}$ shown in Figure \ref{afigs}.

The expression for the total energy can be read from the formulas \eqref{E|}, \eqref{E-}, and \eqref{baryon-v}. We have

\begin{equation}\label{E-s}
\begin{split}
E_{\3q}=&2\g\sqrt{\frac{\s}{\lambda}}
\biggl[
\int^1_0\,\frac{dv}{v^2}\,
\biggl(\ep^{\lambda v^2}\Bigl(1-v^4\,\ep^{2\lambda (1-v^2)}\Bigr)^{-\frac{1}{2}}
-1-v^2\biggr)
+
\int^1_{\sqrt{\frac{\nu}{\lambda}}}\,\frac{dv}{v^2}\,\ep^{\lambda v^2}
\Bigl(1-v^4\,\ep^{2\lambda (1-v^2)}\Bigr)^{-\frac{1}{2}}\biggr] \\
+&\g\sqrt{\frac{\s}{\nu}}
\Bigl(\sqrt{\pi\nu}\,\text{Erfi}(\sqrt{\nu})-\ep^{\nu}+3\kappa\ep^{-2\nu}\Bigr)+C
\,.
\end{split}
\end{equation}
As before, the normalization constant is given by $C=3c$.

A complete description has to include the gluing conditions at the vertex. Such  conditions are given by \eqref{coll-netforce} with $\alpha_1=\alpha_2=\alpha$ and $\alpha_3=\frac{\pi}{2}$. So we have

\begin{equation}\label{collsym-netforce}
2\sin\alpha+1=3\kappa(1+4\nu)\ep^{-3\nu}\,.
\end{equation}
It is convenient to express $\lambda$ as a function of $\nu$. By combining \eqref{collsym-netforce} and \eqref{lambda}, we get 

\begin{equation}\label{sc-lambda}
\lambda(\nu)=-\text{ProductLog}
\biggl(-\frac{2}{\sqrt{3}}\nu\,\ep^{-\nu}
\Bigl(1+2\kappa(1+4\nu)\ep^{-3\nu}-3\kappa^2(1+4\nu)^2\ep^{-6\nu}\Bigr)^{-\tfrac{1}{2}}
\biggr)
\,.
\end{equation}

In summary, the three-quark potential is given by the parametric  
equations \eqref{L-s} and \eqref{E-s}. The parameter $\nu$ takes values in the interval $[0,\nu_\star]$, where $\lambda(\nu_\star)=1$.

The analysis of $E_{\3q}(L)$ in the two limiting cases, long and short distances, is formally similar to that of the previous section. The first step is to learn that $L(\nu)$ is an increasing function on the interval $[0,\nu_\star]$. Moreover, it goes from zero to infinity. Having learned this, we can obtain the short distance behavior by expanding $L(\nu)$ and $E(\nu)$ near $\nu=0$ and then reducing the two 
equations to a single one

\begin{equation}\label{small-L}
E_{\3q}(L)=-\frac{5}{2}\frac{\alpha_{\3q}}{L}+C+2\sigma_0 L+o(L)\,,
\end{equation}
where
\begin{equation}\label{alpha3q-l}
\alpha_{\3q}=-\frac{2}{5}L_0E_0\g\,,
\qquad
\sigma_0=\frac{1}{2L_0}\Bigl(E_1+\frac{L_1}{L_0}E_0\Bigr)\g\,\s
\,.
\end{equation}
The $L_i$'s and $E_i$'s are defined in equations \eqref{L0E1} and \eqref{L0E2}, respectively.

To analyze the long distance behavior, we expand the right hand sides of equations \eqref{L-s} and \eqref{E-s} near $\nu=\nu_\star$, then reduce these equations to 

\begin{equation}\label{large-L}
E(L)=2\sigma L+\text{c}+o(1)\,,
\end{equation}
where
\begin{equation}\label{c-sym}
\text{c}=\g\sqrt{\s}\biggl[
\sqrt{\pi}\text{Erfi}(\sqrt{\nu_\star})
+
2\int_0^1\frac{dv}{v^2}\Bigl(\ep^{v^2}\Bigl(1-v^4\ep^{2(1-v^2)}\Bigr)^{\frac{1}{2}}-1-v^2\Bigr)
+
2\int_{\sqrt{\nu_\star}}^1\frac{dv}{v^2}\ep^{v^2}\Bigl(1-v^4\ep^{2(1-v^2)}\Bigr)^{\frac{1}{2}}
+
\frac{3\kappa\ep^{-2\nu_\star}-\ep^{\nu_\star}}{\sqrt{\nu_\star}}
\biggr]
+C
\,.
\end{equation}
$\text{Erfi}(x)$ denotes the imaginary error function. As before, the constant terms $C$ and $\text{c}$ turn out to be different. 

Now we wish to make some predictions. Otherwise, this will remain as an academic exercise in gauge/string duality. So far there are no lattice data available for the symmetric collinear geometry, therefore we compare the results with the phenomenological ans\"atze and those from the previous section.

\begin{figure}[htbp]
\centering
\includegraphics[width=8.3cm]{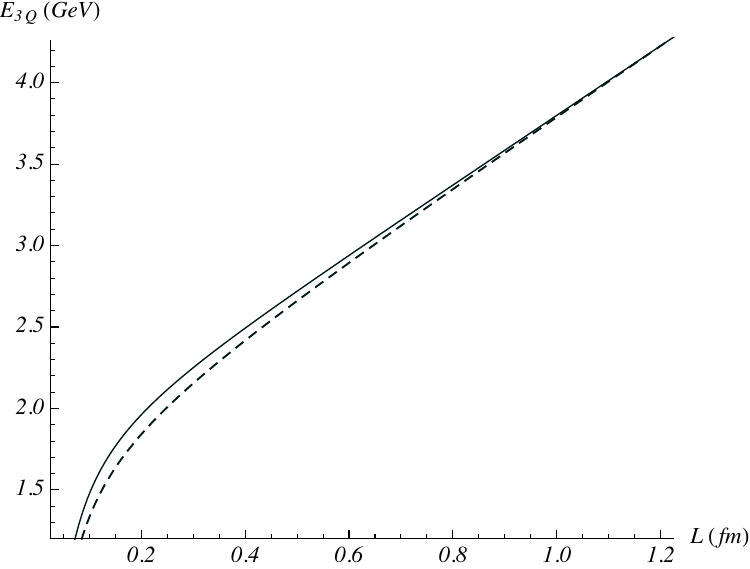}
\caption{\small{The potential $E_{\3q}(L)$ and ansatz \eqref{scEL-Y} (dashed) at $\g=0.176,\,\,s=0.44\,\text{GeV}^2,\,\,\kappa=-0.083$, $C=1.87\,\text{GeV}$, and $c_{\text{\tiny Y}}=1.71\,\text{GeV}$.}}
\label{scpred}
\end{figure}

In Figure \ref{scpred}, 
%________________________  f - 2 __________________________________
%______________________________________________________________
we display the three-quark potential obtained from equations \eqref{L-s} and \eqref{E-s}. Because the phenomenological $\Delta$ and $Y$-laws coincide for the collinear configuration, one could expect that the potential is well described by\footnote{We set $c_{\text{\tiny Y}}=1.71\,\text{GeV}$ to fit on the interval from $0.1\,\text{fm}$ to $1.2\,\text{fm}$.}

\begin{equation}\label{scEL-Y}
E_{\3q}(L)=2\sigma L+c_{\text{\tiny Y}}-\frac{5}{2}\frac{\alpha_{\3q}}{L}
\,.
\end{equation}
However, as seen from Figure \ref{scpred}, this is not the case. The ansatz \eqref{scEL-Y} seems very 
good for short and long distances but not for intermediate ones. The picture is similar to that of 
Figure \ref{et-latticefig}, on the right. In other words, $E_{\3q}$ is a more complicated function of $L$ than the function given by equation \eqref{scEL-Y}.

Let us make some estimates. In units of $\alpha_{\qqb}$ and $\sigma$, we get 

\begin{equation}\label{cstest}
\frac{\alpha_{\3q}}{\alpha_{\qqb}}\approx 0.475
\,,\qquad
\frac{\sigma_0}{\sigma}\approx 0.926
\,.
\end{equation}
The last ratio shows that there is a deviation from the $\Delta$-law defined by \eqref{delta}. 

Looking again at \eqref{cstest} and back to \eqref{etest}, we see that the coefficients in the short distance expansion are not the same. It is a bit surprising but true that for the triangle geometry the coefficients are larger\footnote{We use the subscripts $c$ and $t$ to denote the following geometries: collinear and triangle.}

\begin{equation}\label{ratios}
\frac{\alpha_{\3q}^c}{\alpha_{\3q}^t}\approx 0.960
\,,\qquad
\frac{\sigma_0^c}{\sigma_0^t}\approx 0.920
\,.
\end{equation}
Thus, these expansion coefficients are geometry-dependent.

The long distance expansion is also puzzling. Although the physical string tension is universal (geometry-independent), the constant terms are not. A simple estimate yields

\begin{equation}\label{delta-c}
\text{c}^c-\text{c}^t\approx 0.06\,\text{GeV}
\,.
\end{equation} 
Here we have assumed that the normalization constants $C$ are equal in both cases. 

This is one of the important aspects of the model. The first that comes to mind when thinking about a possible explanation to what we have found is that it has a stringy origin. The point may be that a string junction affects the shape of the three-quark potential not only at long but even at short distances. As a result, the coefficients in the short distance expansion become dependent of the geometry in question. This is another objection to the $\Delta$-law where $\alpha_{\3q}=\oh\alpha_{\qqb}$ for any geometry. Of course, it would be very interesting to see what this means for heavy baryon spectroscopy because neither the $\Delta$ nor $Y$-law takes this effect into account. 

%___________________________________________________________________________________________

\subsection{Generic collinear geometry}

With the experience we have gained, it is straightforward to generalize
formulas such as \eqref{L-s} - \eqref{E-s} and determine the three-quark potential for a generic case of collinear geometry.

Consider the symmetric configuration sketched in Figure \ref{lfig} on the left. To get further, we need to modify it by moving the first quark to a new position, say at $x_1=-l$. Without loss of generality, we may move it closer to the origin such that $l\leq L$. The resulting configuration will be a deformation of the initial one, as shown in Figure \ref{lfig} on the right.

First, we will describe the gluing conditions that is the easier thing to do. From \eqref{coll-netforce}, we immediately obtain 

\begin{equation}\label{gc-forcebalance}
\cos\alpha_1-\cos\alpha_2-\cos\alpha_3=0
\,,\qquad
\sin\alpha_1+\sin\alpha_2+\sin\alpha_3=3\kappa(1+4\nu)\ep^{-3\nu}
\,.
\end{equation}
Here we have abbreviated $\alpha_{\scriptscriptstyle +}{}_i$ to $\alpha_i$.

Next we should mention an important subtlety that arises when one tries to move $Q_1$ closer to $Q_3$. This subtlety is related to a flip of sign in $\alpha_1$, the slope of the first string at the vertex. As a result, the shape of the first string changes from that of Figure \ref{afigs} on the right to that on the left. In the meantime, the other strings keep their shapes.
%______________________________________________________________________

\subsubsection{Configuration with $\alpha_1<0$}

We will now make the discussion more concrete. Consider a small deformation of the symmetric configuration. In that case, we still have $\alpha_1<0$ and $\alpha_2<0$, but with $\alpha_3\not=\frac{\pi}{2}$, as shown in Figure \ref{lfig} on the right.

If we consider the third string, then, in virtue of \eqref{l+}, $\xp$ can be written as an integral. We will thus have

\begin{equation}\label{string3-generic}
\xp=-\cos\alpha_3\,\sqrt{\frac{\nu}{\s}}\int^1_0 dv_3\, v^2_3\, \ep^{\nu(1-v^2_3)}
\Bigl(1-\cos^2{}\hspace{-1mm}\alpha_3\, v^4_3 \,\ep^{2\nu(1-v^2_3)}\Bigr)^{-\frac{1}{2}}
\,.
\end{equation}
Notice that $\xp$ is negative. The reason for this is a force balance at the vertex.

Similarly, for the first and second strings, $\xp +l$ and $L-\xp$ can be expressed in terms of integrals by using equation \eqref{l-}. Combining those with \eqref{string3-generic} leads one to 

\begin{equation}\label{string1-generic}
\begin{split}
l=&\sqrt{\frac{\lambda_1}{\s}}\biggl[
\int^1_0 dv_1\, v^2_1\, \ep^{\lambda_1(1-v^2_1)}
\Bigl(1-v^4_1\,\ep^{2\lambda_1(1-v^2_1)}\Bigr)^{-\frac{1}{2}}+
\int^1_{\sqrt{\frac{\nu}{\lambda_1}}} dv_1\, v^2_1\, \ep^{\lambda_1(1-v^2_1)}
\Bigl(1-v^4_1\,\ep^{2\lambda_1(1-v^2_1)}\Bigr)^{-\frac{1}{2}}
\biggr]\\
&+\cos\alpha_3\,\sqrt{\frac{\nu}{\s}}\int^1_0 dv_3\, v^2_3\, \ep^{\nu(1-v^2_3)}
\Bigl(1-\cos^2{}\hspace{-1mm}\alpha_3\, v^4_3 \,\ep^{2\nu(1-v^2_3)}\Bigr)^{-\frac{1}{2}}
\,
\end{split}
\end{equation}
and 
\begin{equation}\label{string2-generic}
\begin{split}
L=&\sqrt{\frac{\lambda_2}{\s}}\biggl[
\int^1_0 dv_2\, v^2_2\, \ep^{\lambda_2(1-v^2_2)}
\Bigl(1-v^4_2\,\ep^{2\lambda_2(1-v^2_2)}\Bigr)^{-\frac{1}{2}}+
\int^1_{\sqrt{\frac{\nu}{\lambda_2}}} dv_2\, v^2_2\, \ep^{\lambda_2(1-v^2_2)}
\Bigl(1-v^4_2\,\ep^{2\lambda_2(1-v^2_2)}\Bigr)^{-\frac{1}{2}}
\biggr]\\
&-\cos\alpha_3\,\sqrt{\frac{\nu}{\s}}\int^1_0 dv_3\, v^2_3\, \ep^{\nu(1-v^2_3)}
\Bigl(1-\cos^2{}\hspace{-1mm}\alpha_3\, v^4_3 \,\ep^{2\nu(1-v^2_3)}\Bigr)^{-\frac{1}{2}}
\,.
\end{split}
\end{equation}

The energies of the first and second strings can be read from \eqref{E-}, while that of the third from \eqref{E+}. Combining those with the expression \eqref{baryon-v} for the gravitational energy of the vertex, we find the total energy

\begin{equation}\label{E-generic}
\begin{split}
E_{\3q}=&\g\sum_{i=1}^2\sqrt{\frac{\s}{\lambda_i}}
\biggl[
\int^1_0\,\frac{dv_i}{v_i^2}\,
\biggl(\ep^{\lambda_i v^2_i}\Bigl(1-v^4_i\,\ep^{2\lambda_i(1-v^2_i)}\Bigr)^{-\frac{1}{2}}
-1-v^2_i\biggr)
+
\int^1_{\sqrt{\frac{\nu}{\lambda_i}}}\,\frac{dv_i}{v_i^2}\,\ep^{\lambda_i v^2_i}
\Bigl(1-v^4_i\,\ep^{2\lambda_i(1-v^2_i)}\Bigr)^{-\frac{1}{2}}\biggr] \\
+&\g\sqrt{\frac{\s}{\nu}}\biggl[3\kappa\ep^{-2\nu}+
\int^1_0\,\frac{dv_3}{v_3^2}\,\biggl(\ep^{\nu v^2_3}\Bigl(1-\cos^2{}\hspace{-1mm}\alpha_3 
\,v_3^4\,\ep^{2\nu(1-v_3^2)}\Bigr)^{-\frac{1}{2}}-1-v_3^2\biggr)\biggr]
+C
\,.
\end{split}
\end{equation}

First, let us see what happens if $\alpha_3=\frac{\pi}{2}$. In this case, $\xp$ vanishes and $l$ becomes equal to $L$. In addition, the expression \eqref{E-generic} reduces to that of equation \eqref{E-s}. So, the above formulas are consistent with those of subsection C. 

Now, let us try to understand how the potential can be written parametrically as $E_{\3q}=E(\lambda_1,\lambda_2)$, $l=l(\lambda_1,\lambda_2)$, and $L=L(\lambda_1,\lambda_2)$. It is easy to see from \eqref{nu-lambda} that 

\begin{equation}\label{nu-lambda-i}
\cos\alpha_i=\frac{\nu}{\lambda_i}\ep^{\lambda_i-\nu}
\,,
\end{equation} 
with $i=1,2$. After a substitution into the first equation of \eqref{gc-forcebalance}, we find

\begin{equation}\label{cos3}
\cos\alpha_3=\biggl(\frac{\ep^{\lambda_1}}{\lambda_1}-\frac{\ep^{\lambda_2}}{\lambda_2}\biggr)
\nu\,\ep^{-\nu}
\,.
\end{equation}
From this it follows that $\lambda_1$ and $\lambda_2$ must obey $\lambda_1\leq\lambda_2$ if $\lambda_i\in[0,1]$. Now, plugging 
\eqref{nu-lambda-i} and \eqref{cos3} into the second equation of \eqref{gc-forcebalance}, we obtain

\begin{equation}\label{con-generic}
\biggl[1-\frac{\nu^2}{\lambda_1^2}\ep^{2(\lambda_1-\nu)}\biggr]^{\frac{1}{2}}
+\biggl[1-\frac{\nu^2}{\lambda_2^2}\ep^{2(\lambda_2-\nu)}\biggr]^{\frac{1}{2}}
-\biggl[1-\nu^2\ep^{-2\nu}\biggl(\frac{\ep^{\lambda_2}}{\lambda_2}-\frac{\ep^{\lambda_1}}{\lambda_1}\biggr)^2\,\biggr]^{\frac{1}{2}}+3\kappa(1+4\nu)\ep^{-3\nu}=0
\,.
\end{equation}
Unfortunately, we do not know how to explicitly express one parameter as a function of two others. In practice it is convenient to choose the $\lambda$'s as independent parameters and then solve equation \eqref{con-generic} for $\nu$ numerically.

To summarize, the three-quark potential is given in parametrical form by $E_{\3q}=E(\lambda_1,\lambda_2)$, $l=l(\lambda_1,\lambda_2)$, and $L=L(\lambda_1,\lambda_2)$. The parameters take values on the interval $[0,1]$ and obey the inequality $\lambda_1\leq\lambda_2$.

%____________________________________________________________________________________
\subsubsection{Configuration with $\alpha_1=0$}

There is one important situation in which a transition between the two types of profile occurs. This is the case $\alpha_1=0$. Although the calculations are simple, it is useful for the purposes of the present paper to have explicit formulas.

A helpful observation which makes the analysis easy is the following. If $\alpha_1=0$, then $\nu=\lambda_1$, which follows from \eqref{nu-lambda-i}. Hence there is only one independent parameter. 

For $\lambda_1=\nu$ equations \eqref{string1-generic} and \eqref{string2-generic} take the form

\begin{equation}\label{string1-generic0}
l=\sqrt{\frac{\nu}{\s}}\biggl[
\int^1_0 dv_1\, v^2_1\, \ep^{\nu(1-v^2_1)}\Bigl(1-v^4_1\,\ep^{2\nu(1-v^2_1)}\Bigr)^{-\frac{1}{2}}
+
\cos\alpha_3\int^1_0 dv_3\, v^2_3\, \ep^{\nu(1-v^2_3)}
\Bigl(1-\cos^2{}\hspace{-1mm}\alpha_3\, v^4_3 \,\ep^{2\nu(1-v^2_3)}\Bigr)^{-\frac{1}{2}}
\biggr]
\end{equation}
and 
\begin{equation}\label{string2-generic0}
\begin{split}
L=&\sqrt{\frac{\lambda_2}{\s}}\biggl[
\int^1_0 dv_2\, v^2_2\, \ep^{\lambda_2(1-v^2_2)}
\Bigl(1-v^4_2\,\ep^{2\lambda_2(1-v^2_2)}\Bigr)^{-\frac{1}{2}}+
\int^1_{\sqrt{\frac{\nu}{\lambda_2}}} dv_2\, v^2_2\, \ep^{\lambda_2(1-v^2_2)}
\Bigl(1-v^4_2\,\ep^{2\lambda_2(1-v^2_2)}\Bigr)^{-\frac{1}{2}}
\biggr]\\
-&\cos\alpha_3\,\sqrt{\frac{\nu}{\s}}\int^1_0 dv_3\, v^2_3\, \ep^{\nu(1-v^2_3)}
\Bigl(1-\cos^2{}\hspace{-1mm}\alpha_3\, v^4_3 \,\ep^{2\nu(1-v^2_3)}\Bigr)^{-\frac{1}{2}}
\,.
\end{split}
\end{equation}
In the meantime, the expression for the total energy of the configuration becomes

\begin{equation}\label{E-generic0}
\begin{split}
E_{\3q}=&\g\sqrt{\frac{\s}{\nu}}\biggl[3\kappa\ep^{-2\nu}-2+
\int^1_0\frac{dv_1}{v_1^2}
\biggl(\ep^{\nu v^2_1}\Bigl(1-v^4_1\,\ep^{2\nu(1-v^2_1)}\Bigr)^{-\frac{1}{2}}
-1\biggr)
+
\int^1_0\frac{dv_3}{v_3^2}\biggl(\ep^{\nu v^2_3}\Bigl(1-\cos^2{}\hspace{-1mm}\alpha_3 
\,v_3^4\,\ep^{2\nu(1-v_3^2)}\Bigr)^{-\frac{1}{2}}-1\biggr)\biggr]\\
+&\g\sqrt{\frac{\s}{\lambda_2}}\biggl[
\int^1_0\,\frac{dv_2}{v_2^2}\,
\biggl(\ep^{\lambda_2 v^2_2}\Bigl(1-v^4_2\,\ep^{2\lambda_2(1-v^2_2)}\Bigr)^{-\frac{1}{2}}
-1-v^2_2\biggr)
+
\int^1_{\sqrt{\frac{\nu}{\lambda_2}}}\,\frac{dv_2}{v_2^2}\,\ep^{\lambda_2 v^2_2}
\Bigl(1-v^4_2\,\ep^{2\lambda_2(1-v^2_2)}\Bigr)^{-\frac{1}{2}}\biggr] 
+C
\,.
\end{split}
\end{equation}

For completeness, we include the corresponding equations for $\cos\alpha_3$ and $\nu$ in our analysis. A short calculation shows that \eqref{cos3} becomes

\begin{equation}\label{cos30}
\cos\alpha_3=1-\frac{\nu}{\lambda_2}\ep^{\lambda_2-\nu}
\,
\end{equation}
and \eqref{con-generic} becomes 

\begin{equation}\label{con-generic0}
\biggl[1-\frac{\nu^2}{\lambda_2^2}\ep^{2(\lambda_2-\nu)}\biggr]^{\frac{1}{2}}
-\biggl[1-\Bigl(1-\frac{\nu}{\lambda_2}\ep^{\lambda_2-\nu}\Bigr)^2\biggr]^{\frac{1}{2}}+3\kappa(1+4\nu)\ep^{-3\nu}=0
\,.
\end{equation}
Even for this simplified form, we do not know how to explicitly express $\nu$ as a function of $\lambda_2$, but of course it can be done numerically.

Thus, in the case $\alpha_1=0$ the potential can be parametrically written as $E_{\3q}=E(\lambda_2)$, $l=l(\lambda_2)$, and $L=L(\lambda_2)$. The parameter takes values on the interval $[0,1]$. But one important fact about such a configuration is that it occurs only at specific values of $l$ and $L$, as follows from the last two equations.

%____________________________________________________________________________________
\subsubsection{Configuration with $\alpha_1>0$}

Finally, to complete the picture, we need to consider the configurations with $\alpha_1>0$. Such configurations exist for smaller $l$. This can be done by slightly extending what we have described so far. The novelty is that the first string has a shape similar to that of Figure \ref{afigs}, on the left.

First, $\xp +l$ is no longer expressed by \eqref{l-}, but by \eqref{l+}. Since the third string keeps its shape equation \eqref{string3-generic} holds. From this it follows that 

\begin{equation}\label{string1-generic+}
l=\sqrt{\frac{\nu}{\s}}
\sum_{i=1,3}
\int^1_0 dv_i\, v^2_i\, \ep^{\nu(1-v^2_i)}
\cos\alpha_i
\Bigl(1-\cos^2{}\hspace{-1mm}\alpha_i\, v^4_i \,\ep^{2\nu(1-v^2_i)}\Bigr)^{-\frac{1}{2}}
\,.
\end{equation}
At the same time, $L-\xp$ is expressed by \eqref{l-} that makes equation \eqref{string2-generic} true for $\alpha_1>0$ as well.

Second, the energy of the first string is now given by \eqref{E+} rather than by \eqref{E-}. Making this replacement in \eqref{E-generic}, we get the total energy 

\begin{equation}\label{E-generic+}
\begin{split}
E_{\3q}=&\g\sqrt{\frac{\s}{\lambda_2}}
\biggl[
\int^1_0\,\frac{dv_2}{v_2^2}\,
\biggl(\ep^{\lambda_2 v^2_2}\Bigl(1-v^4_2\,\ep^{2\lambda_2(1-v^2_2)}\Bigr)^{-\frac{1}{2}}
-1-v^2_2\biggr)
+
\int^1_{\sqrt{\frac{\nu}{\lambda_2}}}\,\frac{dv_2}{v_2^2}\,\ep^{\lambda_2 v^2_2}
\Bigl(1-v^4_2\,\ep^{2\lambda_2(1-v^2_2)}\Bigr)^{-\frac{1}{2}}\biggr] \\
+&\g\sqrt{\frac{\s}{\nu}}
\biggl[3\kappa\ep^{-2\nu}
+
\sum_{i=1,3} \int^1_0\,\frac{dv_i}{v_i^2}\,\biggl(\ep^{\nu v^2_i}\Bigl(1-\cos^2{}\hspace{-1mm}\alpha_i 
\,v_i^4\,\ep^{2\nu(1-v_i^2)}\Bigr)^{-\frac{1}{2}}-1-v_i^2\biggr)\biggr]
 +C
\,.
\end{split}
\end{equation}

Now let us see how the potential can be written parametrically as $E_{\3q}=E(\nu,\lambda_2)$, $l=l(\nu,\lambda_2)$, and $L=L(\nu,\lambda_2)$. From \eqref{nu-lambda} and \eqref{gc-forcebalance}, it follows that 

\begin{equation}\label{cs1}
\cos\alpha_1=\cos\alpha_3+\frac{\nu}{\lambda_2}\ep^{\lambda_2-\nu}
\,.
\end{equation}
Plugging this into the second equation of \eqref{gc-forcebalance} gives

\begin{equation}\label{cs3}
\biggl[1-\frac{\nu^2}{\lambda_2^2}\ep^{2(\lambda_2-\nu)}\biggr]^{\frac{1}{2}}
-
\biggl[1-\Bigl(\cos\alpha_3+\frac{\nu}{\lambda_2}\ep^{\lambda_2-\nu}\Bigr)^2\biggr]^{\frac{1}{2}}
-
\biggl[1-\cos^2{}\hspace{-1mm}\alpha_3\biggr]^{\frac{1}{2}}
+
3\kappa(1+4\nu)\ep^{-3\nu}=0
\,.
\end{equation}
This equation can be used to solve, at least numerically, for $\cos\alpha_3$ under the given values of $\nu$ and $\lambda_2$.

At this point it is worth noting that at $\alpha_1=0$ all the above formulas reduce to those of the previous section. This fact can be used as a self-consistency check of the results. 

To summarize, the three-quark potential is given in parametric form by $E_{\3q}=E(\nu,\lambda_2)$, $l=l(\nu,\lambda_2)$, and $L=L(\nu,\lambda_2)$. The parameters $\nu$ and $\lambda_2$ take values on the intervals $[0,\nuz ]$ and $[0,1]$, where $\nuz$ is a solution to equation \eqref{con-generic0} at $\lambda_2=1$.

%__________________________________________________________________________________
\subsubsection{What we have learned}

\begin{figure}[htbp]
\includegraphics[width=8.3cm]{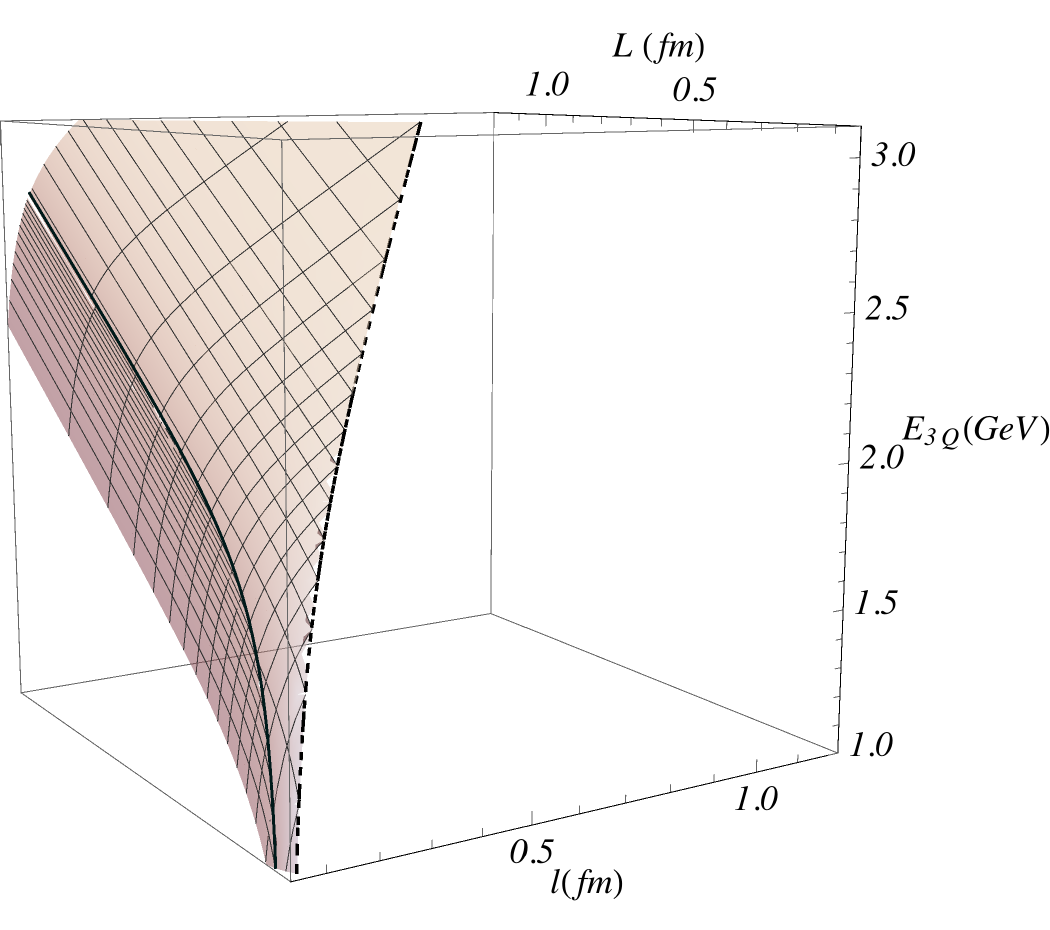}
\caption{{\small The three-quark potential $E_{\3q}(l,L)$ at $\g=0.176,\,\,s=0.44\,{\text GeV}^2,\,\,\kappa=-0.083$ and $C=1.87\,\text{GeV}$. The solid curve indicates where $\alpha_1=0$. The dashed curve represents the potential at $l=L$.}}
\label{3dfig}
\end{figure}

In Figure \ref{3dfig}, we display our result for the potential $E_{\3q}(l,L)$ obtained from the three different types of configurations. For simplicity, we restrict to the 
%________________________  f - 6 __________________________________
%________________________
case $L\geq l$. Since the function $E_{\3q}$ is symmetric under the exchange of $l$ and $L$, one can easily obtain $E_{\3q}$ for the opposite case by a reflection in the line $l=L$.

One lesson to learn from this example is that in the limit $\lambda_2\rightarrow 1$, along the curve given by $\alpha_1=0$, the value of $L$ goes to infinity, while that of $l$ stays bounded. Indeed, it follows from \eqref{string1-generic0} that $l(\lambda_2)$ is an increasing function of $\lambda_2$ whose limiting value at $\lambda_2=1$ is given by 

\begin{equation}\label{max-l}
l_{\text{\tiny max}}=\sqrt{\frac{\nuz}{\s}}
\int^1_0 dv\, 
\biggl[\Bigl(v^{-4}\,\ep^{2\nuz (v^2-1)}-1\Bigr)^{-\frac{1}{2}}
+
\Bigl(\bigl(1-\nuz\ep^{1-\nuz}\bigr)^{-2} v^{-4} \,\ep^{2\nuz (v^2-1)}-1\Bigr)^{-\frac{1}{2}}
\biggr]
\,.
\end{equation}
The same conclusion is true for the configuration with $\alpha_1>0$. The separation $l$ is bounded from above by $l_{\text{\tiny max}}$, while $L$ is not. This can also be seen from Figure \ref{3dfig}.

Let us make a simple estimate of $l_{\text{\tiny max}}$. For the same parameter values as in Figure \ref{3dfig}, the expression \eqref{max-l} yields $l_{\text{\tiny max}}\approx 0.1\,\text{fm}$. This value is compatible with an estimate for a size of the $bb$ diquark obtained from $l\sim 2/m_b$.

Another lesson is that at large separations between the quarks, when both $L$ and $l$ go to infinity, the potential behaves like 

\begin{equation}\label{IR}
E_{\3q}(l,L)=\sigma(L+l)+\text{c}+o(1)\,,
\end{equation}
with $\text{c}$ given by \eqref{c-sym}. To deduce this, we consider the configuration with $\alpha_1<0$ and use the fact that this limiting case corresponds to a small region near the point $(1,1)$ in the parameter space $(\lambda_1,\lambda_2)$. As in subsection C, we expand the right hand sides of \eqref{string1-generic}, \eqref{string2-generic}, and \eqref{E-generic} near $(1,1)$ and then reduce the three equations to the single equation \eqref{IR}.

Besides looking for possible applications to phenomenology, it would be very interesting to confront our predictions with future numerical simulations.

%___________________________________________________________________________________
\section{More on Limiting Cases}
\renewcommand{\theequation}{3.\arabic{equation}}
\setcounter{equation}{0}

In Section II, we took the collinear geometry as the basic example to illustrate the use of the string theory technique in practice. There is more to say if we consider the situation when two quarks get close to each other. In this case, a simplified treatment is to consider a system of two particles, the quark and diquark. The analysis of Section II allows us to shed some light on that situation.

%____________________________________________________________________________________
\subsection{Quark-diquark potential}

It is known that in the limit of large quark masses QCD has a
symmetry which relates hadrons with two heavy quarks to analogous states with one heavy antiquark \cite{wise}. This implies that we should be able to reproduce the heavy quark-antiquark potential from the results of Section II. 

To show this let us consider a collinear configuration with $L$ is fixed and $l\rightarrow 0$.\footnote{One can think of $l$ as a diquark size.} In this case, we should pick the configuration with $\alpha_1>0$. A short analysis shows that in terms of the parameters what we need is to take the limit $\nu\rightarrow 0$ with $\lambda_2$ fixed. 

Letting $\nu\rightarrow 0$ in equations \eqref{cs1} and \eqref{cs3}, we get 

\begin{equation}\label{a13}
\alpha_1=\alpha_3\,,\qquad\text{with}\qquad \alpha_3=\arcsin\frac{1+3\kappa}{2}
\,.
\end{equation}
These formulas mean that in this limit the angle $\alpha_2$ becomes a right angle.

The result of taking $\nu\rightarrow 0$ in \eqref{string1-generic+} is trivial, $l=0$. But in \eqref{string2-generic}, it turns out to be 

\begin{equation}\label{d-L}
L=2\sqrt{\frac{\lambda_2}{\s}}
\int^1_0 dv_2\, v^2_2\, \ep^{\lambda_2(1-v^2_2)}
\Bigl(1-v^4_2\,\ep^{2\lambda_2(1-v^2_2)}\Bigr)^{-\frac{1}{2}}
\,.
\end{equation}
Notice that $L$ is a continuously increasing function of $\lambda_2$. It vanishes at $\lambda_2=0$ and develops a logarithmic singularity at $\lambda_2=1$.

Taking the limit $\nu\rightarrow 0$ in \eqref{E-generic+} requires some care. Because of divergences at $\nu=0$, we need a regulator that renders $E_{\3q}$ finite. The right procedure, which is consistent with what we did before, is to first replace $\nu$ by its lower bound. It is given by $\s\epsilon^2$ as it follows from the lower bound on $r$. Then the regularized expression for $E_{\3q}$ is given by 

\begin{equation}\label{d-ER}
\begin{split}
E_R=&\g\sqrt{\frac{\s}{\lambda_2}}
\biggl[
\int^1_0\,\frac{dv_2}{v_2^2}\,
\biggl(\ep^{\lambda_2 v^2_2}\Bigl(1-v^4_2\,\ep^{2\lambda_2(1-v^2_2)}\Bigr)^{-\frac{1}{2}}
-1-v_2^2\biggr)
+
\int^1_{\sqrt{\frac{\s}{\lambda_2}}\epsilon}\,\frac{dv_2}{v_2^2}\,\ep^{\lambda_2 v^2_2}
\Bigl(1-v^4_2\,\ep^{2\lambda_2(1-v^2_2)}\Bigr)^{-\frac{1}{2}}\biggr] \\
+&\frac{\g}{\epsilon}
\biggl[3\kappa\,\ep^{-2\s\epsilon^2}
+2\int^1_0\,\frac{dv_1}{v_1^2}\,\biggl(\ep^{\s\epsilon^2 v^2_1}\Bigl(1-\cos^2{}\hspace{-1mm}\alpha_1 
\,v_1^4\,\ep^{2\s\epsilon^2(1-v_1^2)}\Bigr)^{-\frac{1}{2}}-1-v_1^2\biggr)\biggr]
 +C
\,.
\end{split}
\end{equation}
A little calculation reveals that the expansion in powers of $\epsilon$ takes the form

\begin{equation}\label{d-ER1}
E_R=\frac{\g}{\epsilon}E_0+E_{\qd}+o(1)
\,,
\end{equation}
where $E_0$ is given by \eqref{lLE-dq3}. Notice that the singular term contains divergences coming from an infinitely large quark mass as well as self-energies of the vertex and strings. It is important that the coefficient $E_0$ does not depend on $\lambda_2$ and, as a consequence, on $L$. This allows one to deal with the divergence in a fashion similar to the standard treatment of power divergences in Wilson loops.

Subtracting the $\frac{1}{\epsilon}$ term and letting $\epsilon=0$, we get a finite result

\begin{equation}\label{d-E}
E_{\qd}=2\g\sqrt{\frac{\s}{\lambda_2}}
\int^1_0\,\frac{dv_2}{v_2^2}\,
\biggl(\ep^{\lambda_2 v^2_2}\Bigl(1-v^4_2\,\ep^{2\lambda_2(1-v^2_2)}\Bigr)^{-\frac{1}{2}}
-1-v_2^2\biggr)+C'
\,.
\end{equation}
Here $C'$ is a normalization constant which is scheme dependent.

Equations \eqref{d-L} and \eqref{d-E} provide a parametric representation of the quark-diquark potential at distances much larger than the diquark size. This parametric representation coincides with that of \cite{az1} for the quark-antiquark potential, as expected \cite{wise}.

%__________________________________________________________________________________________________
\subsection{Quark-quark potential}

\begin{figure}[htbp]
\centering
\includegraphics[width=6.25cm]{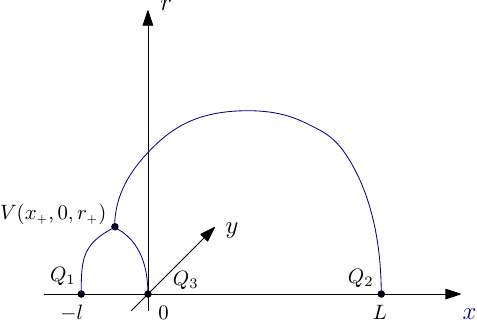}
\caption{\small{A collinear configuration for $L\gg l$.}}
\label{difig}
\end{figure}

We will describe a static potential which represents a part of the inter-quark interaction inside heavy diquarks of small size. As above, we pick the configuration with $\alpha_1>0$. However, we now consider the case when $l$ is small but non-zero, and $L\rightarrow\infty$, as sketched in Figure \ref{difig}.
%________________________  f - 2 __________________________________
%__________________________________________________________________
This means that one quark is placed far away from the others such that its impact on the diquark can be easily assessed. From \eqref{string2-generic} and \eqref{string1-generic+} it follows that in terms of the parameters $\nu$ and $\lambda_2$ we need to consider a small region near the point $(0,1)$ in the parameter space $(\nu,\lambda_2)$. The difference with what we have done in subsection A is that in the expansions in powers of $\nu$ we should keep track of small terms which give rise to a term linear in $l$ in the expression for the total energy.

With the help of the expansions \eqref{lLE-dq}, one readily sees that for small $l$ and large $L$ the 
three-quark potential behaves like

\begin{equation}\label{di-quark}
E_{\3q}(l,L)=-\frac{\alpha_{\qq}}{l}+\sigma_0l+\sigma\Bigl(L+\frac{1}{2}l\,\Bigr)+\text{c}+o(l)\,,
\end{equation}

where 
\begin{equation}\label{di-c}
\alpha_{\qq}=-l_0E_0\g\,,
\quad
\sigma_0=\frac{1}{l_0}\Bigl(E_1+\frac{l_1}{l_0}E_0\Bigr)\g\,\s
\,,
\quad
\text{c}=2\g\sqrt{\s}
\int_0^1\frac{dv}{v^2}
\Bigl(\ep^{v^2}\Bigl(1-v^4\ep^{2(1-v^2)}\Bigr)^{\frac{1}{2}}-1-v^2\Bigr)
\,
+C
\,.
\end{equation}
The $l_i$'s and $E_i$'s are given by equations \eqref{lLE-dq2} and \eqref{lLE-dq3}, respectively. Notice that in \eqref{di-c} the difference between $\text{c}$ and $C$ is written as an integral which is equal to that obtained in the quark-antiquark potential at long distances \cite{hybrids}. This is consistent with the symmetry \cite{wise}.

There is an important subtlety that arises when one tries to extract the quark-quark potential from the above expression. This subtlety is related to the fifth dimension. The point is that in the limit we are considering the $x$-coordinate of the vertex turns out to be $\xp=-l/2$. In other words, the vertex is located between the $Q_1$ and $Q_3$ quarks, as shown in Figure \ref{difig}. In contrast, from the four-dimensional perspective it should be located at $\xp=0$ because in the case of the collinear geometry the vertex (the Fermat point) coincides with $Q_3$.\footnote{Note that in the model we are considering a projection of $V$ on the boundary coincides with the Fermat point only in the infrared limit, when the quarks are far away from each other \cite{a-bar}.} Thus, we have to subtract from $E_{\3q}$ the term proportional to $\sigma$ and a constant term. Such a constant term has to include a contribution from the integral on the right-hand side of the last equation in \eqref{di-c}. These two terms represent the quark-diquark binding energy at large distances. As a result, we get

\begin{equation}\label{diquark}
E_{\qq}(\,l\,)=-\frac{\alpha_{\qq}}{l}+\sigma_0l+C'+o(l)\,,
\end{equation}
with $C'$ a normalization constant. This is the main result of this section.

Having derived the static quark-quark potential, we can make some estimates of $\alpha_{\text{qq}}$ and $\sigma_0$. Using the fitted value of $\kappa$, we get 

\begin{equation}\label{qq-est}
\frac{\alpha_{\qq}}{\alpha_{\qqb}}\approx 0.466 
\,,\qquad
\frac{\sigma_0}{\sigma}\approx 0.785
\,.
\end{equation}
Here we have used $\alpha_{\qqb}$ from the quark-antiquark potential. Thus, our estimates suggest that the ratio $\alpha_{\qq}/\alpha_{\qqb}$ is close to $\tfrac{1}{2}$, and that the effective string tension inside a heavy diquark is approximately 20 percent less than the physical tension. 

In phenomenology\footnote{See, e.g., \cite{Ebert} and references therein.}, the quark-quark potential is usually related to the quark-antiquark potential by Lipkin's rule. This is an ansatz which says that $E_{\qq}=\tfrac{1}{2}E_{\qqb}$ \cite{Lipkin}. Our estimate of $\alpha_{\qq}/\alpha_{\qqb}$ is relatively close to it, whereas that of $\sigma_0/\sigma$ is somewhat larger. 

It is also worth noting that in four dimensions a simple estimate based on a string model yields a larger value for the ratio $\sigma_0/\sigma$. It is $\tfrac{\sqrt{3}}{2}\approx 0.866$ \cite{pdf-diquark}. The caution here is that the reason of decreasing the string tension is geometric and nothing similar happens once three quarks are placed on a straight line.

%___________________________________________________________ section 5
\section{An Example of Hybrid Three-Quark Potential}
\renewcommand{\theequation}{4.\arabic{equation}}
\setcounter{equation}{0}

Here we will describe a special type of hybrid three-quark potentials, namely, the potentials whose gluonic excitations are similar to those in the $\Sigma_u^{-}$ potential for the quark-antiquark. For simplicity, we illustrate only the case when one string is excited. The question we want to address is this: what is the impact of gluonic excitations on the physical picture we found for the ground state?

Consider the configuration of Figure \ref{etfig}. We are going to modify it by exciting one string via gluonic excitations so that the quarks remain at the same positions, i.e., at the vertices of the equilateral triangle. Unfortunately, we know of no efficient way to describe all possible excitations \cite{kuti,bali-pineda} within our model. The only exception is the so called $\Sigma$ excitations which have zero angular momentum. From the five dimensional point of view, such excitations may be modeled by cusps on the Nambu-Goto strings \cite{hybrids}. Importantly, cusps are allowed only in the radial direction that has no impact 
on smoothness of a four-dimensional picture. 

\begin{figure}[htbp]
\centering
\includegraphics[width=5.5cm]{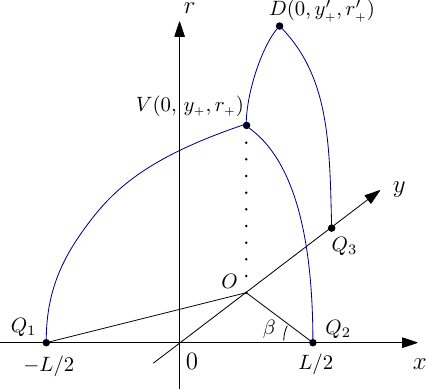}
\caption{\small{A hybrid baryon configuration. The quarks $Q_i$ are placed at the vertices of the equilateral triangle. $V$ and $D$ denote the vertex and defect.}}
\label{hybfig}
\end{figure}

One way to implement this is to insert a local object, called the defect, on a string \cite{hybrids}. Technically, it is quite similar to what we did before for the baryon vertex, with the only difference that two strings join at a defect in the interior as shown in Figure \ref{bfigs}. By inserting the defect on the third string of Figure \ref{etfig}, we construct 
a new configuration shown in Figure \ref{hybfig}. It is an example of a string picture 
%________________________ __________________________________
%__________________________________________________________________
(five-dimensional) of hybrid baryons. Notice that the configuration has a reflection symmetry with respect to the $y$-axis.

Let us begin our analysis with the gluing conditions. As follows from the discussion in the Appendix B, the only condition to be satisfied at the defect is that

\begin{equation}\label{no-force-hybrid-d}
\sin\alpha_3-\kd(1+4\lambda)\ep^{-3\lambda}=0
\,,
\end{equation}
with $\kd=\frac{n}{2\g}$. One can obtain it from equation \eqref{dr} by simply replacing $\nu$ with $\lambda$ and $\alpha_1$ with $\alpha_3$. The conditions at the vertex are a bit more involved since there are two equations for force equilibrium, in the $y$ and $r$-directions. From \eqref{is-netforce} it follows that 

\begin{equation}\label{no-force-hybrid-v}
\cos\alpha'_3=2\cos\alpha_1\sin\beta
\,,\qquad
\sin\alpha'_3-2\sin\alpha_1+3\kappa(1+4\nu)\ep^{-3\nu}=0
\,,
\end{equation}
where $\beta$ is the angle shown in Figure \ref{hybfig}. Notice that $\alpha'_3$ is positive and is related by equation \eqref{am-ap} to $\alpha_3$ so that $\frac{\ep^{\lambda}}{\lambda}\cos\alpha_3=\frac{\ep^{\nu}}{\nu}\cos\alpha'_3$. 

In Section II, we observed that a string shape can change as a quark attached to its endpoint approaches another one. In the example we are analyzing, there is a similar story. A heuristic understanding of this can be obtained without resorting to explicit calculations, as follows. For large $L$, the configuration looks pretty much like that of Figure \ref{etfig} because a local defect has a little impact on very long strings. This implies that $\alpha_1$ is negative. For small $L$, the strings become short, and we might expect the opposite to happen. It does happen when the defect has the dominant effect. In this case, shown in Figure \ref{hybfig}, the third string is stretched along the $r$-axis from the boundary at $r=0$ to a position of the defect at $r=\rp'$, whereas the vertex is located at $r=\rp$ such that $\rp\ll\rp'$. It is obvious that $\alpha_1$ is now positive. This means that a flip of sign in $\alpha_1$ occurs at some value of $L$.

%_________________________________________________________________________________________________
\subsubsection{Configuration with $\alpha_1<0$}

We begin with the case that the non-excited strings have negative slopes at the vertex. Equation \eqref{l-} tells how to write the distance between the points $Q_1$ and $O$ via an integral. It is 

\begin{equation}\label{l1-}
l_1=\sqrt{\frac{\lambda_1}{\s}}\biggl[
\int^1_0 dv_1\, v_1^2\, \ep^{\lambda_1(1-v_1^2)}
\Bigl(1-v_1^4\,\ep^{2\lambda_1(1-v_1^2)}\Bigr)^{-\frac{1}{2}}+
\int^1_
{\sqrt{\frac{\nu}{\lambda_1}}} 
dv_1\, v_1^2\, \ep^{\lambda_1(1-v_1^2)}
\Bigl(1-v_1^4\,\ep^{2\lambda_1(1-v_1^2)}\Bigr)^{-\frac{1}{2}}
\biggr]
\,,
\end{equation}
with $\frac{\ep^{\lambda_1}}{\lambda_1}=\frac{\ep^{\nu}}{\nu}\cos\alpha_1$.

Given \eqref{l+} and \eqref{l++2}, we see that the distance between the point $Q_3$ and $O$ can be represented as a sum of integrals

\begin{equation}\label{l3-}
l_3=\cos\alpha_3\sqrt{\frac{\lambda}{\s}}
\biggl[
\int^1_0 dv_3\, v_3^2\, \ep^{\lambda(1-v_3^2)}
\Bigl(1-\cos^2{}\hspace{-1mm}\alpha_3\, v_3^4 \,\ep^{2\lambda(1-v_3^2)}\Bigr)^{-\frac{1}{2}}
+
\int^1_{\sqrt{\frac{\nu}{\lambda}}} dv_3\, v_3^2\, \ep^{\lambda(1-v_3^2)}
\Bigl(1-\cos^2{}\hspace{-1mm}\alpha_3\, v_3^4 \,\ep^{2\lambda(1-v_3^2)}\Bigr)^{-\frac{1}{2}}
\biggr]
\,.
\end{equation}
Note that in order for the integrals to remain well-defined and finite, the condition must hold that $\lambda<\lam$, where $\lam$ is a solution of equation $\cos\alpha_3=\lambda\,\ep^{1-\lambda}$. In particular, in the limit $\lambda\rightarrow\lam$ one has $l_3\rightarrow\infty$. In this limit, $l_1$ also goes to infinity with $\nu$ going to $\nu_\ast$ which is defined by \eqref{etlambda}. For $\kd=1300$, the numerical calculation gives $\lam\approx 3.29334$. 

It is convenient to use the law of sines to complete the calculation for $L$. In the triangle $OQ_1Q_3$, we have 

\begin{equation}\label{sine}
l_1=\frac{L}{2\cos\beta}=\frac{l_3}{\sqrt{3}\cos\beta-\sin\beta}
\,
\end{equation}
that yields 

\begin{equation}\label{L-hyb-}
L=l_1\cos\beta
\,.
\end{equation}
Coming back to the law of sines, we see that we must also satisfy the constraint

\begin{equation}\label{con-hyb-}
l_3=l_1(\sqrt{3}\cos\beta-\sin\beta)
\,.
\end{equation}

The energies of the first and second strings can be read from \eqref{E-}, while that of the third from \eqref{E+} and \eqref{E++2}. Combining those with \eqref{baryon-v} and \eqref{defect-d} for the gravitational energies of the vertex and defect, we find the total energy\footnote{We use tilde to denote hybrid potentials.}

\begin{equation}\label{E-hyb2-}
\begin{split}
\tilde{E}_{\3q}=&
2\g\sqrt{\frac{\s}{\lambda_1}}
\biggl[
\int^1_0\,\frac{dv_1}{v_1^2}\,
\biggl(
\ep^{\lambda_1 v_1^2}\Bigl(1-v_1^4\,\ep^{2\lambda_1(1-v_1^2)}\Bigr)^{-\frac{1}{2}}-1-v_1^2\biggr)
+
\int^1_{\sqrt{\frac{\nu}{\lambda_1}}}\,\frac{dv_1}{v_1^2}\,
\ep^{\lambda_1v_1^2}\Bigl(1-v_1^4\,\ep^{2\lambda_1(1-v_1^2)}\Bigr)^{-\frac{1}{2}}
\biggr]\\
+&\g\sqrt{\frac{\s}{\lambda}}\biggl[
\int^1_0\,\frac{dv_3}{v_3^2}\,\biggl(\ep^{\lambda v_3^2}\Bigl(1-\cos^2{}\hspace{-1mm}\alpha_3\,v_3^4\,\ep^{2\lambda (1-v_3^2)}\biggr)^{-\frac{1}{2}}-1-v_3^2\Bigr)
+
\int^1_{\sqrt{\frac{\nu}{\lambda}}}\,\frac{dv_3}{v_3^2}\,
\ep^{\lambda v_3^2}\Bigl(1-\cos^2{}\hspace{-1mm}\alpha_3\,v_3^4\,\ep^{2\lambda (1-v_3^2)}\Bigr)^{-\frac{1}{2}}\biggr]
\\
+&
3\kappa\g\sqrt{\frac{\s}{\nu}}\ep^{-2\nu}
+2\kd\g\sqrt{\frac{\s}{\lambda}}\ep^{-2\lambda}
+C
\,.
\end{split}
\end{equation}

Now let us try to understand how the potential can be written parametrically as $\tilde{E}_{\3q}=\tilde{E}(\lambda)$ and $L=L(\lambda)$. This is done in two steps. First, equations \eqref{no-force-hybrid-d} and \eqref{no-force-hybrid-v} imply that the angles can be determined in terms of $\nu$ and $\lambda$. With this result, we have $\tilde{E}_{\3q}=\tilde{E}(\lambda,\nu)$ and $L=L(\lambda,\nu)$. Finally, we use the constraint \eqref{con-hyb-} to write $\nu$ in terms of $\lambda$. Unfortunately, we know of no ways to implement these steps analytically, but only numerically.

To summarize, the hybrid three-quark potential is given in parametrical form by $\tilde{E}_{\3q}=\tilde{E}(\lambda)$, and $L=L(\lambda)$. The parameter takes values on the interval $[\la0,\lam]$, where $\la0$ is defined below.

%____________________________________________________________________________________________
\subsubsection{Configuration with $\alpha_1=0$}

Now we will discuss the case $\alpha_1=0$. As in the example of Section II, at $\alpha_1=0$ a transition between the two types of profile occurs. Since there is only one parameter in the problem, the additional condition should be able to determine its value at the transition point, namely 
$\lambda\vert_{\alpha_1=0}=\la0$.

In this case, it is easy to explicitly show that 

\begin{equation}\label{angles}
\sin\alpha'_3=-3\kappa(1+4\nu)\ep^{-3\nu}
\,,
\qquad
\sin\beta=\oh\Bigl[1-9\kappa^2(1+4\nu)^2\ep^{-6\nu}\Bigr]^\oh
\end{equation}
and, as a consequence, the relation between $\lambda$ and $\nu$ is given by 

\begin{equation}\label{constraint0}
\frac{\ep^{\lambda}}{\lambda}
\Bigl[1-\kd^2(1+4\lambda)^2\ep^{-6\lambda}\Bigr]^{\frac{1}{2}}
=
\frac{\ep^{\nu}}{\nu}
\Bigl[1-9\kappa^2(1+4\nu)^2\ep^{-6\nu}\Bigr]^{\frac{1}{2}}
\,.
\end{equation}

The additional relation is provided by equation \eqref{con-hyb-}. Then, at least in principle, one can find $\nu$ and $\lambda$ by solving the two equations simultaneously at given values of $\kappa$ and $\kd$. In practice, we can do it only numerically. For $\kappa=-0.083$ and $\kd=1300$, we get $\nun\approx 0.99997$ and $\la0\approx 3.29326$. Importantly, we have found that $\la0<\lam$, as expected for consistency. Finally, let us note that the value of $L$ at the transition point turns out to be quite large. For the parameter values quoted above it is about $2.83\,\text{fm}$.

%_______________________________________________________________________________________________
\subsubsection{Configuration with $\alpha_1>0$}

Now let us discuss the case $\alpha_1>0$. We can obtain the hybrid potential by making a simple modification to what we have described so far. The novelty here is that the shape of the non-excited strings changes. This is important for understanding the short distance behavior of $\tilde{E}_{\3q}$.

In this case, $l_1$ is expressed by \eqref{l+} and therefore takes the form

\begin{equation}\label{l1+}
l_1=\cos\alpha_1\,\sqrt{\frac{\nu}{\s}}\int^1_0 dv_1\, v_1^2\, \ep^{\nu(1-v_1^2)}
\Bigl(1-\cos^2{}\hspace{-1mm}\alpha_1\, v_1^4 \,\ep^{2\nu(1-v_1^2)}\Bigr)^{-\frac{1}{2}}
\,.
\end{equation}
In addition, the energies of the non-excited strings are now expressed by \eqref{E+}. Replacing the corresponding terms in \eqref{E-hyb2-}, we obtain the total energy of the configuration 

\begin{equation}\label{E-hyb2+}
\begin{split}
\tilde{E}_{\3q}=&\g\sqrt{\frac{\s}{\nu}}
\biggl[
3\kappa\ep^{-2\nu}
+
2\int^1_0\,\frac{dv_1}{v_1^2}\,
\biggl(
\ep^{\nu v_1^2}\Bigl(1-\cos^2{}\hspace{-1mm}\alpha_1\,
v_1^4\,\ep^{2\nu (1-v_1^2)}\Bigr)^{-\frac{1}{2}}
-1-v_1^2
\biggr)
\biggr]
+
\g\sqrt{\frac{\s}{\lambda}}
\biggl[
2\kd\ep^{-2\lambda}
\\
+&
\int^1_0\,\frac{dv_3}{v_3^2}\,\biggl(\ep^{\lambda v_3^2}\Bigl(1-\cos^2{}\hspace{-1mm}\alpha_3\,v_3^4\,\ep^{2\lambda (1-v_3^2)}\Bigr)^{-\frac{1}{2}}-1-v_3^2\biggr)
+
\int^1_{\sqrt{\frac{\nu}{\lambda}}}\,\frac{dv_3}{v_3^2}\,
\ep^{\lambda v_3^2}\Bigl(1-\cos^2{}\hspace{-1mm}\alpha_3\,v_3^4\,\ep^{2\lambda (1-v_3^2)}\Bigr)^{-\frac{1}{2}}
\biggr]\,+C
\,.
\end{split}
\end{equation}

At this point we can complete the parametric description of the three-quark potential . As before, the law of sines yields the expression for $L$ and the constraint on the number of parameters. Thus, the hybrid three-quark potential is given by $\tilde{E}_{\3q}=\tilde{E}(\lambda)$ and $L=L(\lambda)$, with the parameter taking values in the interval $[\lap,\la0]$. Here $\lap$ is a solution of equation $\sin\alpha_3=1$. Note that in the limit $\lambda\rightarrow\lap$ one has $L\rightarrow 0$ and $\nu\rightarrow 0$. For $\kd=1300$ the numerical calculation gives $\lap\approx 3.27179$ that is consistent with $\lap<\la0$.

%______________________________________________________________________________________________
\subsubsection{What we have learned from this example}

Having derived the expression for the hybrid three-quark potential, we can gain some understanding of what happens to the physical picture we found for the ground state 
if gluonic excitations are of string nature.\footnote{Alternatively, there may be an option of getting hybrid potentials by exciting the baryon vertex.}

It is simplest to begin with the limiting cases: short and long distances $L$. To analyze the short distance behavior of $\tilde{E}_{\3q}$, we consider the limit $\lambda\rightarrow\lap$. First, we expand the right hand sides of \eqref{L-hyb-} and \eqref{E-hyb2+} near $\lambda=\lap$. Next, we reduce these equations to a single equivalent equation

\begin{equation}\label{E-hyb-small}
\tilde{E}_{\3q}(L)=-\frac{\alpha_\qq}{L}+\tilde{C}+\sigma_0 L+ \frac{3}{4} A L^2 +o(L^2)
\,,
\end{equation}
where $\alpha_\qq$ and $\sigma_0$ are given by \eqref{di-c}. In other words, the odd-term coefficients coincide with those of the quark-quark potential. The remaining even-term coefficients are 

\begin{equation}\label{cA}
\tilde{C}=2\g\sqrt{\frac{\s}{\lap}}
\Bigl(\kd\ep^{-2\lap}-\ep^{\lap}+\sqrt{\pi\lap}\text{Erfi}(\sqrt{\lap})\Bigr)+C\,,
\qquad
A=\g\,\s^{\frac{3}{2}}\Bigl(\sqrt{\pi}\text{Erf}(\sqrt{\lap})-2\sqrt{\lap}\ep^{-\lap}\Bigr)^{-1}
\,.
\end{equation}
These coincide with those of the $\Sigma_u^-$ hybrid potential \cite{hybrids}.\footnote{For the constant term, it only makes sense to speak of the scheme independent part $\tilde{C}-C$.} Note that the factor $\tfrac{3}{4}$ is of geometric nature. The excited string is perpendicular to the side of the triangle such that the distance between its endpoints is equal to the height of the triangle $h=\tfrac{\sqrt{3}}{2}L$. 

Equation \eqref{E-hyb-small} has some conceptual interest. Let us first write it as

\begin{equation}\label{hyb-delta}
\tilde{E}_{\3q}(L)=E_\qq(L)+\tilde{E}_{\qqb}(h)+o(L^2)
\,.
\end{equation}
This formula shows that at short distances the hybrid three-quark potential is described by a sum of two-body potentials. Using \eqref{diquark}-\eqref{qq-est} and keeping only the leading terms, we obtain

\begin{equation}\label{hyb-delta1}
\tilde{E}_{\3q}(L)\approx\oh E_{\qqb}(L)+\tilde{E}_{\qqb}(L)
\,.
\end{equation}
This result is notable because it suggests a relation between the hybrid three-quark and quark-antiquark potentials. In the Appendix D, we give a more detailed and precise account of this relation which is in fact a  generalization of the $\Delta$-law.

Another interesting conclusion one can draw from \eqref{E-hyb-small} is that it suggests that the quark-quark potential \eqref{diquark}, which is defined when one starts with $E_{\3q}$ and then takes one quark away from the others, is universal (geometry-independent).\footnote{This issue was raised by Ph.de Forcrand. In an effort to provide more evidence, 
we have carried out the calculation using the same geometry as that of \cite{pdf-diquark}. See the Appendix E for more details.} 

In a similar spirit, we can explore the long distance behavior of $\tilde{E}_{\3q}$. Expanding now the right hand sides near $\lambda=\lam$, we reduce the two equations to a single one

\begin{equation}\label{E-hyb-large}
\tilde{E}_{\3q}(L)=\sqrt{3}\sigma L+\text{c}+\Delta+o(1)\,.
\end{equation}
Here $\text{c}$ is the same constant as that in the expression \eqref{tsigma-c} for the ground state.

The basic fact about equations \eqref{etEL-large} and \eqref{E-hyb-large} is a finite gap between $\tilde{E}_{\3q}$ and $E_{\3q}$ at large $L$. Explicitly, it is given by 

\begin{equation}\label{gap-hybrid}
\Delta=\lim_{L\rightarrow\infty} \tilde{E}_{\3q}(L)-E_{\3q}(L)
=
2\g\sqrt{\s}\biggl[
\frac{\kd}{\sqrt{\lam}}\ep^{-2\lam}
+
\int^{\sqrt{\lam}}_1\frac{dv}{v^2}\ep^{v^2}\Bigl(1-v^4\ep^{2(1-v^2)}\Bigr)^{\frac{1}{2}}
\biggr]\,.
\end{equation}
The result agrees with the calculation of \cite{hybrids} for the hybrid quark-antiquark potentials. This suggests that the gap $\Delta$, like the physical string tension $\sigma$, is universal (geometry independent). It is one of the main results of this paper.

\begin{figure}[htbp]
\centering
\includegraphics[width=8.25cm]{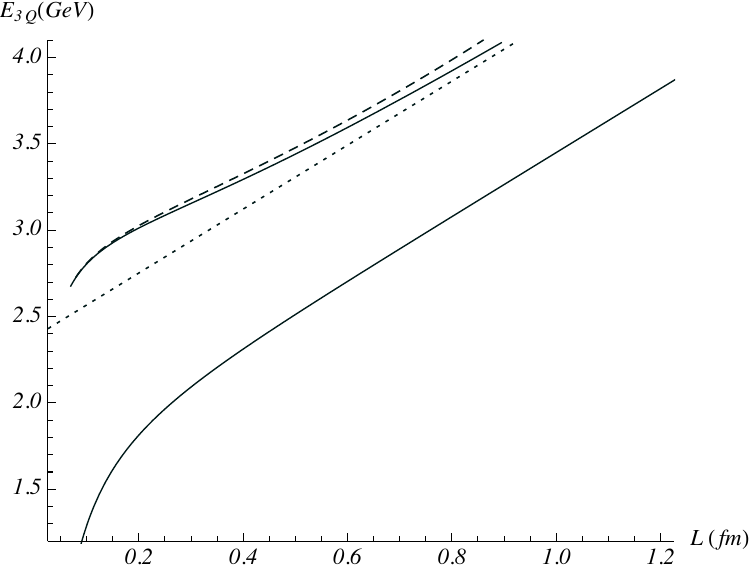}
\caption{\small{The static three-quark potential and its hybrid excitation obtained on an equilateral triangle of side length $L$. The dashed curve corresponds to the $\Delta$-law \eqref{E-hyb-small} and the dotted one to the Y-law \eqref{E-hyb-large}.
}}
\label{hyb}
\end{figure}

Now a question arises: What happens in between? In Figure \ref{hyb}, we display our results for the three-quark potentials. 

%________________________  f - hyb __________________________________
%__________________________________________________________________
\noindent Here, in addition to the parameters of Figure \ref{et-latticefig}, there is the parameter $\kd$. 
We set $\kd=1300$ by fitting the gap to $\Delta=0.78\,\text{GeV}$.\footnote{This value was determined from the spectra of closed string excitations \cite{teper-dub}. It is quite close to that determined from the hybrid potentials \cite{kuti,hybrids}. This fact also suggests the universality of the gap.} Notice that it deviates from the value $\kd=2000$ used in \cite{hybrids}. This might seem strange, but it is easily explained. The point is that $A$ is a slow varying function of $\kd$. Indeed, a numerical calculation shows that $A(2000)/A(1300)\approx 0.99$. This is the reason for the deviation if $\kd$ is fitted to $A$. 

For large $L$, our model reveals the Y-law, as it follows from the slope in \eqref{E-hyb-large}. In contrast, the $\Delta$-law \eqref{E-hyb-small} yields the smaller slope equal to that of Section II. It is worth noting that $\tilde{E}_{\3q}$ approaches the linear behavior at larger separations than $E_{\3q}$ does. The linear approximation is accurate enough for $\tilde{E}_{\3q}$ when $L\geq 1\,\text{fm}$ compared to $L\geq 0.6\,\text{fm}$ for $E_{\3q}$ \cite{a-Nq}. For small $L$, the $\Delta$-law is a good approximation to $\tilde{E}_{\3q}$. However, the deviation increases with 
$L$, as seen from Figure \ref{hyb}. Thus, our model predicts that for hybrid potentials the physical picture remains essentially the same: it incorporates two-body interactions at short distances and a genuine three-body interaction at longer ones.

We conclude our discussion of this example by making a couple of remarks. First, one can see from Figure \ref{hyb} that the function $\tilde{E}_{\3q}(L)$ has an inflection point at $L=L_i$. The expression \eqref{E-hyb-small} allows one to make an estimate of $L_i$. It is $L_i\approx 0.34\,\text{fm}$. Interestingly, in \eqref{E-hyb-small} the $1/L$ and $L^2$ terms cancel each other at $L=L_i$. Second, we have considered the case when only one string is excited. The generalization to the case when two or three strings are excited is straightforward, but technical and more elaborate.

%_________________________________________________________ section 5
\section{Concluding Comments}
\renewcommand{\theequation}{5.\arabic{equation}}
\setcounter{equation}{0}

The purpose of this paper has been to make predictions based on the success of previous work \cite{az1,a-Nq,hybrids}. The motivation was to gain some understanding of how baryons are put together from quarks.

The model we are pursuing has led to a number of interesting conclusions:

(1) The usual view on the three-quark potential is that the potential can be well approximated by the $\Delta$-law \eqref{delta} at short distances and by the Y-law \eqref{Y} at long ones.\footnote{Interestingly, a similar pattern also occurs in three dimensions \cite{3d}. Here, it was found that a L\"uscher-like correction, which is subleading to a constant term in the long-distance expansion of the three-quark potential, is geometry-dependent \cite{3dL}.} Mathematically, the point is that it is described by a complicated function whose asymptotic behavior is given by those laws. We found that a further refinement of this view should take into account that in \eqref{delta} and \eqref{Y} the coefficients $\alpha_{\3q}$ and $\text{c}$ are not universal but dependent of a geometry of sources (quarks). Translating into mathematical language, the coefficients depend on angular variables. Physically, this is a signature of genuine three-body interactions, even at short distances. 

(2) The given example of a hybrid three-quark potential suggests that one may regard the $\Delta$ and $Y$-laws as good approximations to the hybrid potentials as well. 

(3) There are at least two possible ways to cross check our findings. One way is by performing lattice simulations and another is by multi-loop calculations in perturbative QCD \cite{bram}.  

(4) The model we are developing is an effective string theory based on the Nambu-Goto formalism in a curved space. Therefore it has some limitations. In particular, an issue of the L\"uscher-like correction remains unclear. Another unclear issue is that a force of attraction occurs between the vertex and the boundary. What could be the reason for this? $m$ is a result of a resummation of infinitely many terms ($\alpha'$ corrections) in the five-brane action. Is it negative because the brane tension is negative, and if so, does it lead to instability?\footnote{This is not so obvious because a baryonic configuration is not a single brane, but a bound state, and the value we need is quite small. There has been an extensive discussion of negative tension branes but in other contexts \cite{negativeT}.} These questions have no obvious answers. The Green-Schwarz formalism, already developed for strings on $\text{AdS}_5\times\mathbf{S}^5$ \cite{rrm}, seems more appropriate to address these issues. However, this is still far from having been implemented. It is even not clear whether the ansatz \eqref{metric10}, which is a deformation of $\text{AdS}_5\times\mathbf{S}^5$, is a solution of the supergravity equations. And, if so, what role do the other background fields play in QCD? But what is clear is that finding the way to the string description of QCD is a challenging and difficult problem, and along the way we will see and experience a lot of things \cite{polyakov}. In the meantime, lattice gauge theory and effective string 
models will remain the main tools of investigation.

%__________________________________________________________________
\begin{acknowledgments}
We would like to thank P.de Forcrand, A. Leonidov, H. Suganuma, and P. Weisz for helpful discussions. We also wish to thank P.de Forcrand and P. Weisz for reading the manuscript and for making valuable suggestions. This work was supported in part by Russian Science Foundation grant 16-12-10151. Finally, we would like to thank the Arnold Sommerfeld Center for Theoretical Physics and CERN Theory Division for the warm hospitality. 
\end{acknowledgments}

%_____________________________________________________________________________________
\appendix
\section{A static Nambu-Goto string with fixed endpoints}
%\label{notation}
\renewcommand{\theequation}{A.\arabic{equation}}
\setcounter{equation}{0}

In this appendix, we will discuss a static Nambu-Goto string in the curved geometry \eqref{metric10}. The results provide the grounds for building multi-string configurations.

The Nambu-Goto action is given by

\begin{equation}\label{NG}
S=\frac{1}{2\pi\alpha'}\int_0^1 d\sigma\int_0^T d\tau\,\sqrt{\gamma}
\,,
\end{equation}
with $\gamma$ an induced metric on the string world-sheet (with Euclidean signature). Consider a string stretched between the two fixed points $P(\xm,\rm)$ and $B(\xp,\rp)$ in the $xr$-plane such that $\xp\geq\xm$ and $\rp\geq\rm$. This implies the following boundary conditions 

\begin{equation}\label{string-bc}
x(0)=\xm\,,\quad x(1)=\xp\,,\quad r(0)=\rm\,,\quad r(1)=\rp
\,.
\end{equation}

In static gauge, $t=\tau$ and $x=a\sigma +b$, the action takes the form

\begin{equation}\label{string-NG}
S=T\g\int_{\xm}^{\xp} dx\,w(r)\sqrt{1+(\partial_x r)^2}
\,,
\qquad
\text{with}
\qquad
w(r)=\frac{\ep^{\s r^2}}{r^2}
\,.
\end{equation}
For convenience, we use the shorthand notation $\g=\frac{R^2}{2\pi\alpha'}$ and $\partial_x r=\frac{\partial r}{\partial x}$. Since the integrand in \eqref{string-NG} does not depend explicitly on $x$, the Euler-Lagrange equation has the first integral

\begin{equation}\label{I}
I=\frac{w(r)}{\sqrt{1+(\partial_x r)^2}}\,.
\end{equation}
At the endpoints, $I$ can be written as

\begin{equation}\label{I-PB}
I=w(r_{\scriptscriptstyle \pm})\cos\alpha_{\scriptscriptstyle\pm}\,,
\end{equation}
where $\tan\alpha_{\scriptscriptstyle \pm}=\partial_xr\vert_{x=x_{\scriptscriptstyle \pm}}$ and 
$\alpha_{\scriptscriptstyle\pm}\in[-\frac{\pi}{2},\frac{\pi}{2}]$.

\begin{figure}[htbp]
\centering
\includegraphics[width=5.25cm]{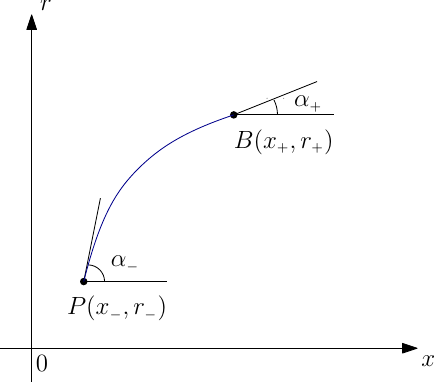}
\hspace{4cm}
\includegraphics[width=5.35cm]{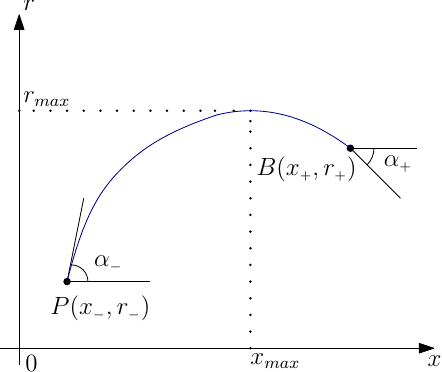}
\caption{{\small A string stretched between two points. $\alpha_{\scriptscriptstyle\pm}$ denote the tangent angles at these points. We assume that $\rp<1/\sqrt{\s}$. Left: The case $\ap>0$. Right: The case $\ap<0$.}}
\label{afigs}
\end{figure}

In general, $\ap$ can be positive or negative. If $\ap>0$, then a function $r(x)$ describing a string profile is increasing on the interval $[\xm,\xp]$. If $\ap<0$, then the situation is a little bit more complicated. $r(x)$ is increasing on the interval $[\xm,x_{\text{\tiny max}}]$ and decreasing on the interval $[x_{\text{\tiny max}},\xp ]$ such that it has a maximum at $x=x_{\text{\tiny max}}$. Examples for both cases are sketched in Figure \ref{afigs}. 
%________________________  f - A __________________________________
%________________________
First we describe the case of positive $\ap$.

\renewcommand \thesubsubsection {\arabic{subsubsection}}
\subsubsection{ The case $\ap\geq 0$}

We begin with the case in which $P$ lies on the boundary, that means $\rm=0$. After expressing $I$ in terms of $\ap$ and $\rp$, we get a differential equation $w(\rp)\cos\ap=w(r)/\sqrt{1+(\partial_x r)^2}$ that can be integrated over the variables $x$ and $r$. So, we have

\begin{equation}\label{l+}
\ell=\xp-\xm=\cos\ap\sqrt{\frac{\nup}{\s}}\int^1_0 dv\, v^2\, \ep^{\nup (1-v^2)}
\Bigl(1-\cos^2{}\hspace{-1mm}\ap v^4 \,\ep^{2\nup(1-v^2)}\Bigr)^{-\frac{1}{2}}
\,,
\end{equation}
where $\nu_{\scriptscriptstyle{\pm}}=\s r_{\scriptscriptstyle{\pm}}^2$. This integral is well-defined for $\nup<1$, while for larger values of $\nup$ it becomes ill-defined unless $\nup\,\ep^{1-\nup}\geq \cos\ap$. 

Having derived the expression for the string length along the $x$-axis, we can give a formal recipe for computing the energy of the string. First, we reduce the integral over $x$ in $S$ to that over $r$. This is easily done by using the first integral. Since the integral is divergent at $r=0$, we regularize it by imposing a cutoff $\epsilon$ such that $r\geq\epsilon$. Finally, the regularized expression is given by 

\begin{equation}\label{e+}
E_{R}=\frac{S_R}{T}=\g\sqrt{\frac{\s}{\nup}}\int^1_{\sqrt{\tfrac{\s}{\nup}}\epsilon}\,\frac{dv}{v^2}\,\ep^{\nup v^2}\Bigl(1-\cos^2{}\hspace{-1mm}\ap\,v^4\,\ep^{2\nup(1-v^2)}\Bigr)^{-\frac{1}{2}}
\,.
\end{equation}
In the limit as $\epsilon\rightarrow 0$, it behaves like 

\begin{equation}\label{E+R}
E_R=\frac{\g}{\epsilon}+E+O(\epsilon)\,.
\end{equation}
Subtracting the $\tfrac{1}{\epsilon}$ term and letting $\epsilon=0$, we get a finite result

\begin{equation}\label{E+}
E=\g\sqrt{\frac{\s}{\nup}}
\int^1_0\,\frac{dv}{v^2}\,\biggl(\ep^{\nup v^2}
\Bigl(1-\cos^2{}\hspace{-1mm}\ap\,v^4\,\ep^{2\nup (1-v^2)}\Bigr)^{-\frac{1}{2}}-1-v^2\biggr)
+c
\,,
\end{equation}
with $c$ a normalization constant. 

If we set $\ap=\frac{\pi}{2}$, then the string becomes straight. In that case, we obtain

\begin{equation}\label{E|}
\ell=0\,,
\qquad
E=\g\sqrt{\frac{\s}{\nup}}
\Bigl(\sqrt{\pi\nup}\,\text{Erfi}(\sqrt{\nup})-\ep^{\nup}\Bigr)+c\,,
\end{equation}
which is the form in which it is written in \cite{pol-loop}. $\text{Erfi}(x)$ is the imaginary error function.

It is straightforward to extend this analysis to the case where the endpoint $P$ is not
on the boundary but in the interior such that $\rm >\epsilon$. As before, we express $I$ in terms of $\ap$ and $\rp$ and then integrate the differential equation. After doing so, we find

\begin{equation}\label{l++2}
\ell
=\cos\ap\,\sqrt{\frac{\nup}{\s}}\int_{\sqrt{\frac{\num}{\nup}}}^1 dv\, v^2\, \ep^{\nup (1-v^2)}\Bigl(1-\cos^2{}\hspace{-1mm}\ap\, v^4 \,\ep^{2\nup(1-v^2)}\Bigr)^{-\frac{1}{2}}
\,.
\end{equation}
The integral is well-defined for $\nu_{\scriptscriptstyle{\pm}}>1$. It becomes ill-defined for smaller values
unless $\cos\ap\leq \nup\,\ep^{1-\nup}$ for $\num<1$ 
and $\cos\ap\leq \frac{\nup}{\num}\ep^{\num-\nup}$ for $\nu_{\scriptscriptstyle{\pm}}<1$ .

Then, we compute the energy
 
\begin{equation}\label{E++2}
E=\g\sqrt{\frac{\s}{\nup}}
\int_{\sqrt{\frac{\num}{\nup}}}^1\,\frac{dv}{v^2}\,
\ep^{\nup v^2}\Bigl(1-\cos^2{}\hspace{-1mm}\ap\,
v^4\,\ep^{2\nup (1-v^2)}\Bigr)^{-\frac{1}{2}}
\,.
\end{equation}
In that case, there is no need for regularization.

We could of course have defined the first integral in terms of $\am$ and $\rm$. This gives an equivalent result

\begin{equation}\label{l++}
\ell=
\cos\am\,\sqrt{\frac{\num}{\s}}\int^{\sqrt{\frac{\nup}{\num}}}_1 dv\, v^2\, \ep^{\num(1-v^2)}
\Bigl(1-\cos^2{}\hspace{-1mm}\am\, v^4 \,\ep^{2\num(1-v^2)}\Bigr)^{-\frac{1}{2}}
\,,
\end{equation}
\begin{equation}\label{E++}
E=\g\sqrt{\frac{\s}{\num}}
\int^{\sqrt{\frac{\nup}{\num}}}_1\,\frac{dv}{v^2}\,\ep^{\num v^2}\Bigl(1-\cos^2{}\hspace{-1mm}\am\,v^4\,\ep^{2\num(1-v^2)}\Bigr)^{-\frac{1}{2}}
\,.
\end{equation}
It is easy to see that a rescaling $v\rightarrow\sqrt{\frac{\nup}{\num}}\,v$ reduces these expressions to those in \eqref{l++2} and \eqref{E++2}. Another useful formula is

\begin{equation}\label{am-ap}
\cos\am \frac{\ep^{\num}}{\num}=\cos\ap\frac{\ep^{\nup}}{\nup}
\end{equation}
which follows from \eqref{I-PB}.

\subsubsection{The case $\ap\leq 0$}

What we have done so far generalizes straightforwardly to the case $\ap\leq 0$. An important point, which applies to all expressions written below, is that there are two contributions: one comes from the interval $[\xm,x_{\text{\tiny max}}]$ on which $r(x)$ is increasing and the other from the interval $[x_{\text{\tiny max}},\xp]$ on which $r(x)$ is decreasing (see Figure \ref{afigs}). Since the case $\rm\not=0$ is not so important for our applications, we put the string endpoint $P$ on the boundary. 

If we define the first integral at $r=r_{\text{\tiny max}}$ such that $I=w(r_{\text{\tiny max}})$, and then integrate the differential equation \eqref{I} over both intervals, we get

\begin{equation}\label{l-}
\ell=\sqrt{\frac{\lambda}{\s}}\biggl[
\int^1_0 dv\, v^2\, \ep^{\lambda(1-v^2)}
\Bigl(1-v^4\,\ep^{2\lambda(1-v^2)}\Bigr)^{-\frac{1}{2}}+
\int^1_
{\sqrt{\frac{\nup}{\lambda}}} 
dv\, v^2\, \ep^{\lambda(1-v^2)}
\Bigl(1-v^4\,\ep^{2\lambda(1-v^2)}\Bigr)^{-\frac{1}{2}}
\biggr]
\,,
\end{equation}
with $\lambda=\s r_{\text{\tiny max}}^2$ and $\nup <\lambda$. The integrals are well-defined if $\lambda$ takes values in the interval $[0,1]$. 

The alternative to what we have done is to directly use \eqref{l+} and \eqref{l++2}. The first term is obtained by setting $\ap=0$ in \eqref{l+} and then replacing $\nup$ by $\lambda$. By combining this with $\num\rightarrow\nup$ in \eqref{l++2}, one reaches the desired result.

It is worth noting that $\lambda$, $\nup$, and $\ap$ are not independent. From \eqref{I} it follows that 

\begin{equation}\label{nu-lambda}
\frac{\ep^{\lambda}}{\lambda}=\frac{\ep^{\nup}}{\nup}\cos\ap
\,.
\end{equation}
Moreover, $\lambda$, as a function of $\nup$ and $\ap$, can be written in terms of the ProductLog function as 

\begin{equation}\label{lambda}
\lambda=-\text{ProductLog}(-\nup\,\ep^{-\nup}/\cos\ap)\,,
\end{equation}
where $\text{ProductLog}(z)$ is the principal solution for $w$ in $z=w\,\ep^w$ \cite{wolf}.

As in our previous case, the energy of the string can computed by first replacing the integral over $x$ in $S$ to that over $r$ and then imposing the short-distance cutoff on $r$. A calculation along the above lines gives

\begin{equation}\label{e-}
E_{R}=\g\sqrt{\frac{\s}{\lambda}}
\biggl[
\int^1_{\sqrt{\tfrac{\s}{\lambda}}\epsilon}\,\frac{dv}{v^2}\,\ep^{\lambda v^2}
\Bigl(1-v^4\,\ep^{2\lambda (1-v^2)}\Bigr)^{-\frac{1}{2}}
+
\int^1_{\sqrt{\frac{\nup}{\lambda}}}\,\frac{dv}{v^2}\,\ep^{\lambda v^2}
\Bigl(1-v^4\,\ep^{2\lambda (1-v^2)}\Bigr)^{-\frac{1}{2}}\biggr] 
\,.
\end{equation}
To obtain from $E_R$ a finite result, we need to subtract the $\frac{1}{\epsilon}$ term 
and then let $\epsilon=0$. Finally, we are left with

\begin{equation}\label{E-}
E=
\g\sqrt{\frac{\s}{\lambda}}
\biggl[
\int^1_0\,\frac{dv}{v^2}\,
\biggl(\ep^{\lambda v^2}\Bigl(1-v^4\,\ep^{2\lambda (1-v^2)}\Bigr)^{-\frac{1}{2}}
-1-v^2\biggr)
+
\int^1_{\sqrt{\frac{\nup}{\lambda}}}\,\frac{dv}{v^2}\,\ep^{\lambda v^2}
\Bigl(1-v^4\,\ep^{2\lambda (1-v^2)}\Bigr)^{-\frac{1}{2}}\biggr] 
+c
\,,
\end{equation}
where $c$ is the normalization constant. By essentially the same arguments that we have given for $\ell$, the expression for $E$ can be obtained directly from \eqref{E+} and \eqref{E++2}.

We conclude this discussion with a couple of remarks. 

First of all, in the case $\ap=0$ the expressions \eqref{l-} and \eqref{E-} are equal to those obtained in the previous section. Clearly, in this situation, the string profiles sketched 
in Figure \ref{afigs} coincide with each other.

Although the case $\ap=\frac{\pi}{2}$ is trivial, we have presented it above 
for completeness reasons, the case $\ap=-\frac{\pi}{2}$ needs a special care. The key point is that the string endpoint $B$ approaches the boundary as $\ap\rightarrow -\frac{\pi}{2}$. Clearly, if $I$ is finite, then $w(\rp)$ has to be infinite. This means that $\nup$ goes to $0$ as $\ap$ goes to $-\frac{\pi}{2}$. There is no difficulty with \eqref{l-} because both integrals are convergent. Thus, we have 

\begin{equation}\label{r2}
\ell=2\sqrt{\frac{\lambda}{\s}}
\int_0^1 dv\,v^2\ep^{\lambda(1-v^2)}
\Bigl(1-v^4\ep^{2\lambda(1-v^2)}\Bigr)^{-\oh}
\,.
\end{equation}
However, in \eqref{E-} the second integral becomes divergent at $v=0$. We cut off the integral by placing a lower bound $\sqrt{\frac{\s}{\lambda}}\epsilon$. This is consistent with what we did before. Indeed, a lower bound $r\geq \epsilon$ implies a lower bound for $\nup$ of the form 
$\nup\geq \s\epsilon^2$. As a result, the regularized expression is simply

\begin{equation}\label{E2r}
E_R= \g\sqrt{\frac{\s}{\lambda}}
\biggl[
\int^1_0\,\frac{dv}{v^2}\,
\biggl(\ep^{\lambda v^2}\Bigl(1-v^4\,\ep^{2\lambda (1-v^2)}\Bigr)^{-\frac{1}{2}}
-1-v^2\biggr)
+
\int^1_{\sqrt{\frac{\s}{\lambda}}\epsilon}\,\frac{dv}{v^2}\,\ep^{\lambda v^2}
\Bigl(1-v^4\,\ep^{2\lambda (1-v^2)}\Bigr)^{-\frac{1}{2}}\biggr] 
+c
\,.
\end{equation}
It diverges for $\epsilon\rightarrow 0$ as $\frac{\g}{\epsilon}$. After subtracting the 
divergence, we get a finite result

\begin{equation}\label{E2}
E=2\g\sqrt{\frac{\s}{\lambda}}
\int_0^1 \frac{dv}{v^2}\biggl(\ep^{\lambda v^2}
\Bigl(1-v^4\ep^{2\lambda (1-v^2)}\Bigr)^{-\oh}-1-v^2\biggr)
\,\,+ 2c
\,.
\end{equation}
Here $\lambda$ is a parameter taking values in $[0,1]$. The normalization constant is equal to $2c$ because of the double subtraction of $\frac{1}{\epsilon}$. Note that the expressions \eqref{r2} and \eqref{E2} coincide with those obtained in \cite{az1} for a string stretched between points on the boundary, as expected.

%_______________________________ appendix B ____________________________________
\section{Gluing conditions}
%\label{notation}
\renewcommand{\theequation}{B.\arabic{equation}}
\setcounter{equation}{0}

The string solutions discussed in the Appendix A provide the basic blocks for building multi-string configurations. What is also needed are certain gluing conditions for such blocks. The goal here is to describe those conditions. 

From the physical viewpoint, the gluing conditions arise as follows. Strings meet at common junctions such as baryon vertices or defects. Clearly, a static baryon configuration must obey the condition that a net force vanishes at any junction. It is straightforward to translate this into a mathematical language. Extremizing the total action of a system of Nambu-Goto strings and junctions with respect to a location of the $i$-junction, one gets a force balance condition at that location\footnote{See also \cite{liu}.}. To illustrate how this works in practice, we follow \cite{hybrids, a-bar} and consider two basic examples: two strings meeting at a defect and three strings meeting at a baryon vertex.

\setcounter{subsubsection}{0}
\subsubsection{Two strings meeting at a defect}

It is simplest to begin with two strings meeting at a defect. Such a defect results in a cusp formation in the $r$-direction that represents a kind of string excitation \cite{hybrids}.

\begin{figure}[tbp]
\centering
\includegraphics[width=5.5cm]{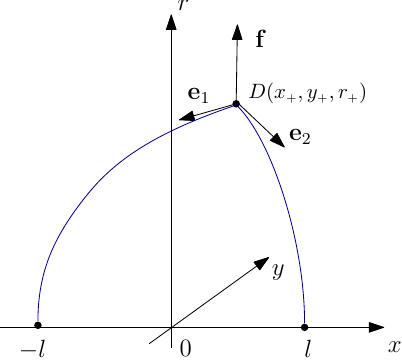}
\hspace{3cm}
\includegraphics[width=5.5cm]{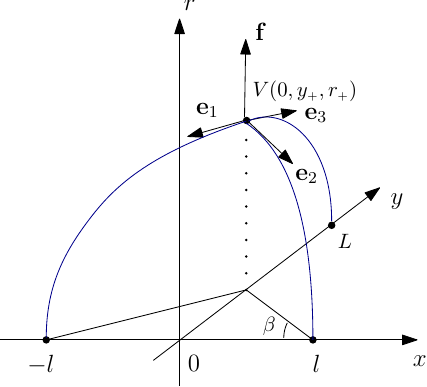}
\caption{{\small String meeting at junctions. The gravitational force acting on junctions is directed in the upward vertical direction. Left: Two strings meeting at a defect placed at $D$.  Right: Three strings meeting at a baryon vertex placed at $V$.}}
\label{bfigs}
\end{figure}

We take two Nambu-Goto strings beginning at the heavy quark sources on the boundary and ending on the defect in the interior, as shown in Figure \ref{bfigs} on the left. In this case, the total action of the  
%________________________  f - B __________________________________
%________________________
configuration reads 

\begin{equation}\label{d-totalaction}
S=\sum_{i=1}^2S_i\,+S_{\text{def}}
\,,
\end{equation}
where $S_i$ is the action of the $i$-string and $S_{\text def}$ is that of the defect. It is given by \eqref{defect-d}.

Since we are interested in a static configuration, it is convenient to choose gauge conditions

\begin{equation}\label{d-gauge}
t_i(\tau_i)=\tau_i
\,,\qquad
x_i(\sigma_i)=a_i\sigma_i+b_i
\,,
\end{equation}
with $(\tau_i,\sigma_i)$ the world-sheet coordinates. Then the action of the $i$-string takes the form

\begin{equation}\label{dNG}
S_i=T\g \int_0^1 d\sigma_i\,w\sqrt{a_i^2+y'_i{}^2+r'_i{}^2}
\,,
\end{equation}
where a prime denotes a derivative with respect to $\sigma_i$.

For the configuration shown in Figure \ref{bfigs}, the boundary conditions on the fields are given by 

\begin{equation}\label{d-boundary}
x_i(0)=\mp\ell\,, 
\qquad
x_i(1)=\xp\,,
\qquad
y_i(0)=0\,,
\qquad
y_i(1)=y_{\scriptscriptstyle +}\,,
\qquad
r_i(0)=0\,,
\qquad
r_i(1)=\rp\,.
\end{equation}
Using these conditions, one can find the coefficients $a_i$ and $b_i$. The result is $a_i=x_+\pm\ell$ and $b_i=\mp\ell$.

A useful observation is that the Euler-Lagrange equations derived from $S_i$ have the first integrals

\begin{equation}\label{dI}
I_i=\frac{w_i}{\sqrt{a_i^2+y'_i{}^2+r'_i{}^2}}\,,
\qquad
P_i=y'_i
\,.
\end{equation}
that allows one to determine the $y_i$'s. From \eqref{d-boundary} it follows that $y_i(\sigma_i)=y_{\scriptscriptstyle +}\sigma_i$.

It is now straightforward to extremize the total action with respect to the location of the defect. A short calculation gives

\begin{equation}\label{deq}
a_1I_1+a_2I_2=0\,,
\qquad
(I_1+I_2)y_{\scriptscriptstyle +}=0\,,
\qquad
r'_1(1)I_1+r'_2(1)I_2+\g^{-1}\partial_{\rp}{\cal V}(\rp)=0
\,,
\end{equation}
with ${\cal V}(r)=n\frac{\ep^{-2\s r^2}}{r}$, as follows from \eqref{defect-d}. We have defined the $I_i$'s at $\sigma_i=1$. 

From the second equation it follows that $y_{\scriptscriptstyle +}=0$. In other words, the static configuration lies in the $xr$-plane. In that case, the first equation reduces to $\cos\ap{}_1=\cos\ap{}_2$, with $\ap{}_i$ the tangent angles at $D$. Its solution is simply $\ap{}_1=\ap{}_2$.\footnote{Note that the parameter used in \cite{hybrids} is given by $k=\tan^2\ap$.} Of course, all of this may be understood using the symmetry analysis alone. The underlying symmetry is reflection about the $y$-axis.

The remaining equation now reads

\begin{equation}\label{dr}
\sin\ap{}_1-\kd(1+4\nup)\ep^{-3\nup}=0
\,, 
\end{equation}
where $\kappa_{\text{d}}=\frac{n}{2\g}$. 

What we have derived by extremizing the total action with respect to the location of the defect
is nothing else but a force balance condition encoding the requirement that in a static baryon configuration the net force must vanish at a defect. It is instructive to see explicitly why it is so. If we take $\mathbf{e}_{i}$ to be a unit tangent vector at a string endpoint which represents a force exerted by the $i$ string on the defect, as shown in Figure \ref{bfigs}, then a gravitation force exerted on the defect is to be normalized as $\mathbf{f}=(0,0,\frac{n}{\g}(1+4\nup)\ep^{-3\nup})$.\footnote{We denote vectors by boldface letters.} Because there is no net force acting on an object in equilibrium, one has 

\begin{equation}\label{d-netforce}
\sum_{i=1}^2\mathbf{e}_{i}+\mathbf{f}=0
\,.
\end{equation}

A solution to the equation for the $y$-component is trivial. It is $\ep_y{}_i=0$. Thus the configuration is effectively two-dimensional that allows one to write $\mathbf{e}_1=(-\cos\ap{}_1,0,-\sin\ap{}_1)$ and $\mathbf{e}_2=(\cos\ap{}_2,0,-\sin\ap{}_2)$. Then the remaining two equations give $\cos\ap{}_1=\cos\ap{}_2$ and $\sin\ap{}_1=\kappa_{\text d}(1+4\nup)\ep^{-3\nup}$. This is exactly what we have been looking for.

The approach based on the force balance condition is effective, and gives a practical recipe for writing the gluing conditions at junctions. It can straightforwardly be extended to include a baryon vertex.

\subsubsection{Three strings meeting at baryon vertex}

At this stage, we consider three Nambu-Goto strings beginning at the heavy quark sources on the boundary and ending on the baryon vertex in the interior, as shown in Figure \ref{bfigs}, on the right. Because the action for the baryon vertex \eqref{baryon-v} is of the form \eqref{defect-d}, with $n$ replaced by $m$, we may write

\begin{equation}\label{v-netforce}
\sum_{i=1}^3\mathbf{e}_{i}+\mathbf{f}=0
\,,
\end{equation}
where $\mathbf{f}=(0,0,\frac{m}{\g}(1+4\nup)\ep^{-3\nup})$.

Let us now consider some examples. 

The first example is a configuration in which the quarks are placed at the vertices of an equilateral triangle, as shown in Figure \ref{etfig}. The symmetry of the problem is the group $D_3$. Hence the projection of $V$ onto the $xy$ plane is simply a center of the triangle and all the tangent angles at $V$ are equal to one another $\ap{}_1=\ap{}_2=\ap{}_3$. In terms of components, the first two of \eqref{v-netforce} are trivially satisfied, while the third gives 

\begin{equation}\label{et-netforce}
\sin\ap{}_1=\kappa(1+4\nup)\ep^{-3\nup}
\,,
\end{equation}
with $\kappa=\frac{m}{3\g}$. We have used that $\ep_r{}_i=-\sin\ap{}_i$. It is noteworthy that $\ap$ may be positive or negative depending on the direction of the gravitational force. 
Alternatively, one could derive the gluing condition \eqref{et-netforce} by varying the total action with respect to the position of the baryon vertex but this way is longer. 

The second example is a collinear configuration of the quarks as shown in Figure \ref{lfig}, on the right. Like in the case of two quarks, the equation for the $y$-component of \eqref{v-netforce} yields a trivial solution $\ep_y{}_i=0$ which means that the configuration is two-dimensional. Hence one can choose 

\begin{equation}
\mathbf{e}_1=(-\cos\ap{}_1,0,-\sin\ap{}_1)\,,
\quad 
\mathbf{e}_2=(\cos\ap{}_2,0,-\sin\ap{}_2)\,,
\quad 
\mathbf{e}_3=(-\cos\ap{}_3,0,-\sin\ap{}_3)\,.
\end{equation}

Then the equations for the remaining components become

\begin{equation}\label{coll-netforce}
\cos\ap{}_1-\cos\ap{}_2-\cos\ap{}_3=0
\,,
\qquad
\sum_{i=1}^3\sin\ap{}_i=3\kappa (1+4\nup)e^{-3\nup}\,.
\end{equation}

Our final example is more complicated but not by much. Consider a configuration where the quarks are placed at the vertices of an isosceles triangle as shown in Figure \ref{bfigs}. Since the configuration has a reflection symmetry with respect to the 
$y$-axis, we have $\ap{}_1=\ap{}_2$. In this case, the unit vectors $\mathbf{e}_i$ are defined as 

\begin{equation}
\begin{split}
\mathbf{e}_1=&(-\cos\beta\cos\ap{}_1,-\sin\beta\cos\ap{}_1,-\sin\ap{}_1)\,,
\\
\mathbf{e}_2=&(\cos\beta\cos\ap{}_1,-\sin\beta\cos\ap{}_1,-\sin\ap{}_1)\,,
\\
\mathbf{e}_3=&(0,\cos\ap{}_3,-\sin\ap{}_3)\,,
\end{split}
\end{equation}
where $\beta$ is the angle shown in Figure \ref{bfigs}.

The equation for the $x$-component of \eqref{v-netforce} is trivial, while the two others are 
given by 

\begin{equation}\label{is-netforce}
\cos\ap{}_3=2\sin\beta\cos\ap{}_1\,,
\qquad
2\sin\ap{}_1+\sin\ap{}_3=3\kappa (1+4\nup)e^{-3\nup}
\,.
\end{equation}

The above examples hopefully make the general idea clear. One writes force balance conditions at junctions, and interprets those as the gluing conditions for strings. Moreover, this is a practical way which works for all multi-string configurations with an arbitrary number of quark sources.

%_____________________________________________________________________________________
\section{Limiting cases}
%\label{notation}
\renewcommand{\theequation}{C.\arabic{equation}}
\setcounter{equation}{0}
This appendix collects together some of the formulas which are used to analyze the short distance behavior of the potential in Sections II-IV.

In the case of the equilateral triangle configuration, we consider equations \eqref{etL} and \eqref{etE}. A simple analysis reveals that small $L$ corresponds to small $\nu$. However, rather than using expansions in powers of $\nu$, it is somewhat more convenient to expand $L$ and $E$ in terms of $\lambda$. This may be done using the relation 

\begin{equation}\label{et-nu}
\nu=\rho^{\frac{1}{2}}\lambda+(1-2\kappa^2-\rho^{\frac{1}{2}})\lambda^2+O(\lambda^3)
\,
\end{equation}
which follows from \eqref{etlambda}. Here $\rho=1-\kappa^2$. As a consequence, one can show that the behavior of $L$ and $E_{\3q}$ near $\lambda=0$ is given by  

\begin{equation}\label{etLE-small}
L(\lambda)=\sqrt{\frac{\lambda}{\s}}\Bigl(L_0+L_1\lambda+O(\lambda^2)\Bigr)\,,
\qquad
E_{\3q}(\lambda)=\g\sqrt{\frac{\s}{\lambda}}\Bigl(E_0+E_1\lambda+O(\lambda^2)\Bigr)+C\,,
\end{equation}
with 
\begin{gather}
L_0=\frac{\sqrt{3}}{4}{\cal B}\bigl(\kappa^2;\tfrac{1}{2},\tfrac{3}{4}\bigr)
\,,\qquad
L_1=\frac{\sqrt{3}}{4}
\Bigl({\cal B}\bigl(\kappa^2;-\tfrac{1}{2},\tfrac{3}{4}\bigr)-
{\cal B}\bigl(\kappa^2;-\tfrac{1}{2},\tfrac{5}{4}\bigr)-
2\rho^{\frac{1}{4}}\vert\kappa\vert^{-1}(1-2\kappa^2-\rho^{\frac{1}{2}}\bigr)
\Bigr)
\,,\label{l0e1}\\
E_0=3\kappa\rho^{-\frac{1}{4}}+
\frac{3}{4}{\cal B}\bigl(\kappa^2;\tfrac{1}{2},-\tfrac{1}{4}\bigr)
\,,\qquad
E_1=-6\kappa\rho^{\frac{1}{4}}+
\frac{3}{4}{\cal B}\bigl(\kappa^2;\tfrac{1}{2},\tfrac{1}{4}\bigr)+\sqrt{3}L_1
\,.
\label{l0e2}
\end{gather}
Here ${\cal B}(z;a,b)=B(a,b)+B(z;a,b)$ and $B(z;a,b)$ is the incomplete beta function.

In the case of the symmetric collinear configuration, we consider equations \eqref{L-s} and \eqref{E-s}. Here it is important to know that if $\nu$ goes to zero, then the integral \eqref{L-s} also goes to zero and, as a consequence, small $L$ corresponds to small $\nu$. As before, we expand $L$ and $E_{\3q}$ in terms of $\lambda$. In this case, \eqref{et-nu} is replaced by 

\begin{equation}\label{cs-nu}
\nu=\zeta\lambda+\Bigl(\zeta^2-\zeta+\frac{3}{4}\kappa(1-3\kappa)\Bigr)\lambda^2+O(\lambda^3)
\,,
\end{equation}
with $\zeta=\frac{\sqrt{3}}{2}(1+2\kappa-3\kappa^2)^{\frac{1}{2}}$, as it follows from \eqref{sc-lambda}. The behavior of $L$ and $E_{\3q}$ near $\lambda=0$ is given by \eqref{etLE-small} but with the coefficients replaced by

\begin{gather}
L_0=\frac{1}{4}{\cal B}\bigl(1-\zeta^2;\tfrac{1}{2},\tfrac{3}{4}\bigr)
\,,\qquad
L_1=\frac{1}{4}\Bigl(
{\cal B}\bigl(1-\zeta^2;-\tfrac{1}{2},\tfrac{3}{4}\bigr)
-
{\cal B}\bigl(1-\zeta^2;-\tfrac{1}{2},\tfrac{5}{4}\bigr)
+
4\frac{1-\zeta}{1-3\kappa}\zeta^{\frac{3}{2}}
-
3\kappa\zeta^{\frac{1}{2}}\Bigr)
\,,\label{L0E1}
\\
E_0=(3\kappa-1)\zeta^{-\frac{1}{2}}+\frac{1}{2}{\cal B}\bigl(1-\zeta^2;\tfrac{1}{2},-\tfrac{1}{4}\bigr)
\,,\qquad
E_1=(1-6\kappa)\zeta^{\frac{1}{2}}+\frac{1}{2}{\cal B}\bigl(1-\zeta^2;-\tfrac{1}{2},\tfrac{1}{4}\bigr)+2L_1
\,.
\label{L0E2}
\end{gather}

In the case of the collinear configuration with large $L$ and small $l$, we consider equations \eqref{string2-generic}, \eqref{string1-generic+}, and \eqref{E-generic+}. One important difference between this case and the ones we have discussed so far is that there are now two parameters. We have to analyze the behavior of the corresponding equations in a small region near the point $(0,1)$ in a two-dimensional parameter space $(\nu,\lambda_2)$. A first step in doing so is to choose a path from the point $(\nu,\lambda_2)$ to $(0,1)$. There is an ambiguity in choosing the path. For definiteness, we choose the following path $\nu=\varepsilon$ and $1-\sqrt{\lambda_2}=\varepsilon$, with $\varepsilon$ a small parameter. This enables us to have expansions in terms of $\varepsilon$ and also control path-dependent terms. Then the expansions are

\begin{equation}\label{lLE-dq}
\begin{split}
l=&\sqrt{\frac{\nu}{\s}}\Bigl(l_0+l_1\nu\Bigr)+o(\varepsilon^{\frac{3}{2}})\,,
\qquad
L=\frac{1}{\sqrt{\s}}\Bigl(-\ln\bigl(1-\sqrt{\lambda}\,\bigr)+{\cal L}_0\Bigr)
-\frac{1}{2}\sqrt{\frac{\nu}{\s}}\,l_0 
+o(\varepsilon^{\frac{1}{2}})\,,\\
E_{\3q}=&\g\sqrt{\s}\Bigl(-\ep\ln\bigl(1-\sqrt{\lambda}\,\bigr)+{\cal E}_0\Bigr)
+C
+\g\sqrt{\frac{\s}{\nu}}\Bigl(E_0+E_1\nu\Bigr)
+o(\varepsilon^{\frac{1}{2}})\,,
\end{split}
\end{equation}
where 
\begin{gather}
l_0=\frac{1}{2}\xi^{-\frac{1}{2}}B\bigl(\xi^2;\tfrac{3}{4},\tfrac{1}{2}\bigr)\,,
\qquad
l_1=\frac{1}{2}\xi^{-\frac{3}{2}}\Bigl(
 \Bigl(2\xi+\frac{3}{4}\frac{\kappa-1}{\xi}\Bigr)B\bigl(\xi^2;\tfrac{3}{4},-\tfrac{1}{2}\bigr)
 -
B\bigl(\xi^2;\tfrac{5}{4},-\tfrac{1}{2}\bigr)\Bigr)\,,
\label{lLE-dq2}\\
E_0=1+3\kappa+\frac{1}{2}\xi^{\frac{1}{2}}B\bigl(\xi^2;-\tfrac{1}{4},\tfrac{1}{2}\bigr)\,,
\qquad
E_1=-1-6\kappa+\frac{1}{2}\xi^{-\frac{1}{2}}B\bigl(\xi^2;\tfrac{1}{4},\tfrac{1}{2}\bigr)
+\xi\,l_1 \,,
\label{lLE-dq3}
\end{gather}
and 
\begin{equation}\label{lLE-dq4}
{\cal L}_0=
\int^1_0 dv\biggl(2\Bigl(v^{-4}\ep^{2(v^2-1)}-1\Bigr)^{-\frac{1}{2}}
-\frac{1}{1-v}\biggr)\,,
\quad
{\cal E}_0=2\int^1_0 \frac{dv}{v^2}\biggl(\ep^{v^2}\Bigl(1-v^4\ep^{2(1-v^2)}\Bigr)^{-\frac{1}{2}}
-1-v^2-\frac{\ep\, v^2}{2(1-v)}\biggr)
\,.
\end{equation}
Here $\xi=\frac{\sqrt{3}}{2}(1-2\kappa-3\kappa^2)^{\frac{1}{2}}$. In \eqref{lLE-dq} we indicate the dependence on $\nu$ and $\lambda$ rather than on $\varepsilon$ when no ambiguity arises. Notice that the integrals in \eqref{lLE-dq4} can be evaluated numerically with the result ${\cal L}_0\approx -0.710$ and ${\cal E}_0\approx-3.431$.

%_____________________________________________________________________________________
\section{More on the $\Delta$-law}
\renewcommand{\theequation}{D.\arabic{equation}}
\setcounter{equation}{0}
The goal here is to discuss the relation between the three and two-quark potentials

\begin{equation}\label{1/2}
E_{\3q}(\mathbf{x}_1,\mathbf{x}_2,\mathbf{x}_3)=\oh\Bigl(
E_{\qqb}(\vert\mathbf{x}_{12}\vert)
+
E_{\qqb}(\vert\mathbf{x}_{23}\vert)
+
E_{\qqb}(\vert\mathbf{x}_{13}\vert)
\Bigr)
\,
\end{equation}
which is called the $\Delta$-law.\footnote{It is sometimes also called $\frac{1}{2}$ rule.} This relation is valid at least to order $\alpha_s^2$ in perturbation theory \cite{review} and, therefore, at very small separations between the quarks. However, it is no longer valid at larger separations, where it is assumed to be replaced by the following inequality\footnote{Interestingly, a similar inequality was discussed before in the context of few-nucleon systems \cite{richard}.}

\begin{equation}\label{1/2-1}
E_{\3q}(\mathbf{x}_1,\mathbf{x}_2,\mathbf{x}_3)\geq\oh\Bigl(
E_{\qqb}(\vert\mathbf{x}_{12}\vert)
+
E_{\qqb}(\vert\mathbf{x}_{23}\vert)
+
E_{\qqb}(\vert\mathbf{x}_{13}\vert)
\Bigr)
\,
\end{equation}
that is consistent with the $Y$-law for $E_{\3q}(\mathbf{x}_1,\mathbf{x}_2,\mathbf{x}_3)$.

A natural question to ask at this point is what kind of relations exist between hybrid potentials? Some relations can be deduced by assuming a factorization in the large $T$ limit such that  

\begin{equation}\label{1/2-2}
\langle W_{\3q}(\mathbf{x}_1,\mathbf{x}_2,\mathbf{x}_3,T)\rangle =
\prod_{i>j}^3
\langle
W_{\qqb}(\vert\mathbf{x}_{ij}\vert,T)
\rangle^\oh
\,.
\end{equation}
This assumption seems reasonable for small $\vert\mathbf{x}_{ij}\vert$.

The simplest case to consider is an equilateral triangle configuration. For such a geometry, the factorization equation becomes 

\begin{equation}\label{1/2-3}
\langle W_{\3q}(L,T)\rangle=
\langle
W_{\qqb}(L,T)
\rangle^{\frac{3}{2}}
\,,
\end{equation}
with $L$ a triangle's side length. In the limit $T\rightarrow\infty$, the expectation values of the Wilson loops are given by 

\begin{equation}\label{series}
\langle W_{\3q}(L,T)\rangle=\sum_{n=0}^\infty B_n(L)\,\ep^{-E_{\3q}^{(n)}(L)T}
\,,\qquad
\langle W_{\qqb}(L,T)\rangle=\sum_{n=0}^\infty M_n(L)\,\ep^{-E_{\qqb}^{(n)}(L)T}
\,.
\end{equation}

For our purposes, it is sufficient to consider the first three energy levels. In that case, equation \eqref{1/2-3} can be easily solved. First, we assume that $E_{\3q}^{(n+1)}>E_{\3q}^{(n)}$ and $E_{\qqb}^{(n+1)}>E_{\qqb}^{(n)}$ hold in the range of small $L$. In other words, the energy levels are separated from each other. In addition, we assume that gaps are subject to the constraint $\Delta E_{\qqb}^{(2)}<2\Delta E_{\qqb}^{(1)}$, where $\Delta E_{\qqb}^{(i)}=E_{\qqb}^{(i)}-E_{\qqb}^{(0)}$. The factorization equation is then simply

\begin{equation}\label{series2}
B_0\,\ep^{-E_{\3q}^{(0)}T}+
B_1\,\ep^{-E_{\3q}^{(1)}T}+B_2\,\ep^{-E_{\3q}^{(2)}T}
=M_0^{\frac{3}{2}}\,\ep^{-\frac{3}{2}E_{\qqb}^{(0)}T}
\biggl(
1+
\frac{3M_1}{2M_0}
\ep^{-\Delta E_{\qqb}^{(1)}T}
+
\frac{3M_2}{2M_0}\,\ep^{-\Delta E_{\qqb}^{(2)}T}
\biggl)
\,,
\end{equation}
modulo terms decaying faster as $T\rightarrow\infty$. Equating order by order the two sides of equation gives

\begin{equation}\label{Delta-hyb}
E_{\3q}^{(0)}(L)=\frac{3}{2}E_{\qqb}^{(0)}(L)\,,
\qquad
E_{\3q}^{(i)}(L)=\oh E_{\qqb}^{(0)}(L)+E_{\qqb}^{(i)}(L)
\,.
\end{equation}
The first equation is nothing but the $\Delta$-law, whereas the second is its  generalization to excited levels. In practice \cite{kuti}, the assumptions we made are valid if $L<0.7\,\text{fm}$. In that case, $E_{\qqb}^{(1)}$ corresponds to the $\Pi_u$ hybrid potential and $E_{\qqb}^{(2)}$ to the $\Sigma_u^-$ one. 

At this point, we should mention that for large $L$ the $Y$-law suggests the inequalities 

\begin{equation}\label{Delta-hyb2}
E_{\3q}^{(0)}(L)\geq\frac{3}{2}E_{\qqb}^{(0)}(L)\,,
\qquad
E_{\3q}^{(i)}(L)\geq\oh E_{\qqb}^{(0)}(L)+E_{\qqb}^{(i)}(L)
\,.
\end{equation}
These, in turn, suggest the following inequalities between meson and baryon masses 

\begin{equation}\label{masses}
2 M (\text{\small QQQ}) \geq 3 M(\text{\small Q}\bar{\text{\small Q}} )\,,
\qquad
2 M(\text{\small QQQg})  \geq M (\text{\small Q}\bar{\text{\small Q}}) +2M(\text{\small Q}\bar{\text{\small Q}}\text{\small g})
\,.
\end{equation}
While the first is known in the literature \cite{richard}, the second is new. It involves the masses of hybrids.

\begin{figure}[t!]
\centering
\includegraphics[width=8cm]{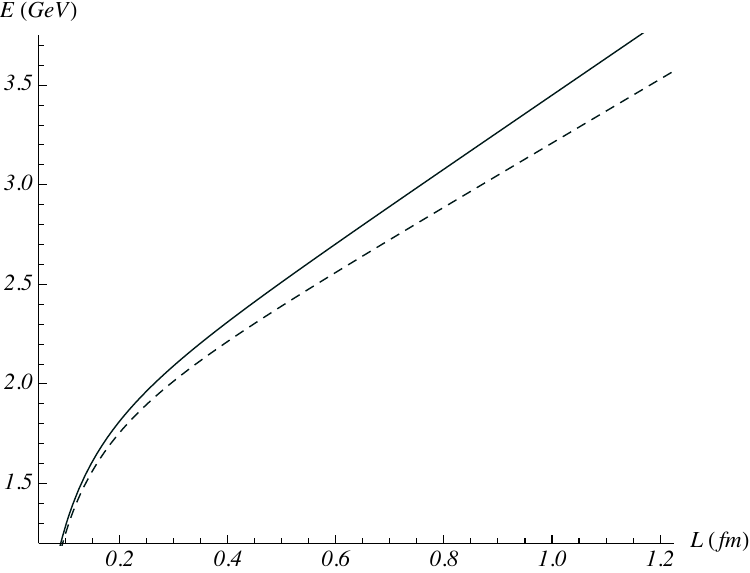}
\hspace{1.5cm}
\includegraphics[width=8cm]{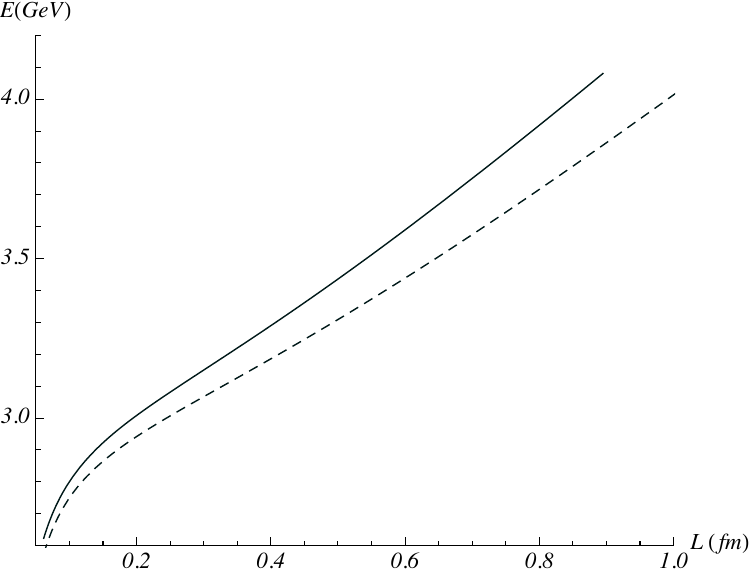}
\caption{{\small Left: $E_{\3q}$ (solid curve) and $\frac{3}{2}E_{\qqb}$ (dashed curve). Right: $\tilde{E}_{\3q}$ (solid curve) and $\frac{1}{2}E_{\qqb}+\tilde{E}_{\qqb}$ (dashed curve). In both cases we set $\g=0.176$, $\s=0.44\,\text{GeV}^2$, $\kappa=-0.083$, $\kd=1300$, and $C=1.87\,\text{GeV}$. $E_{\qqb}$ and $\tilde{E}_{\qqb}$ are from \cite{az1} and \cite{hybrids}, respectively.}}
\label{inq}
\end{figure}

As a final comment, we note that the model we are considering allows us to check two of these inequalities numerically. Figure \ref{inq} shows the results for the ground states and excited states corresponding to the $\Sigma$ excitation.
%________________________  f - B __________________________________
%________________________
Obviously, the three-quark potentials can be approximated by a set of the two-quark potentials only at short distances.

%_____________________________________________________________________________________
\section{Another calculation of the quark-quark potential}
\renewcommand{\theequation}{E.\arabic{equation}}
\setcounter{equation}{0}
Here, we will carry out a calculation of the static quark-quark potential using the same geometry as that of \cite{pdf-diquark}. The result will be equivalent to what we had before. This provides more evidence that the potential \eqref{diquark} is geometry-independent.

\begin{figure}[htbp]
\centering
\includegraphics[width=4.9cm]{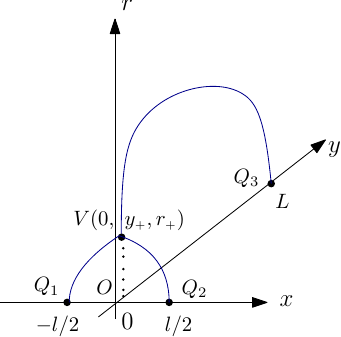}
\caption{\small{A baryon configuration for $L\gg l$. The quarks $Q_i$ are placed at the vertices of the isosceles triangle. $V$ is the vertex and $O$ is its projection on the $y$-axis.}}
\label{ifig}
\end{figure}

Consider a geometry in which the quarks are at the vertices of an isosceles triangle. We are interested in the limiting case when the base length $l$ is much smaller than the triangle's height $L$, as shown in Figure \ref{ifig}. Since the configuration has a reflection symmetry with respect 
%________________________ __________________________________
%__________________________________________________________________
to the $y$-axis, the side strings have an identical profile and the projection of $V$ on the $xy$-plane lies on the $y$-axis. Notice that $\alpha_1$ is positive. This can be understood as follows. Consider a configuration similar to the one shown in Figure \ref{difig} such that $Q_3$ lies on the $x$-axis. Obviously, one can smoothly move $Q_3$ to the $y$-axis along a circle of radius $L$. In the limit $l\rightarrow 0$ and $L\rightarrow\infty$, this may not lead to a radical change of the string's profiles.

Given this, we can use the general formula \eqref{l+} to write the distance between the points $Q_1$ and $O$ via an integral

\begin{equation}\label{Tl1}
l_1=\cos\alpha_1\,\sqrt{\frac{\nu}{\s}}\int^1_0 dv_1\, v_1^2\, \ep^{\nu(1-v_1^2)}
\Bigl(1-\cos^2{}\hspace{-1mm}\alpha_1\, v_1^4 \,\ep^{2\nu(1-v_1^2)}\Bigr)^{-\frac{1}{2}}
\,.
\end{equation}
In a similar fashion, equation \eqref{l-} tells us how to do so for the distance between the points $Q_3$ and $O$. Here we have

\begin{equation}\label{Tl3}
l_3=\sqrt{\frac{\lambda}{\s}}\biggl[
\int^1_0 dv_3\, v^2_3\, \ep^{\lambda(1-v^2_3)}
\Bigl(1-v^4_3\,\ep^{2\lambda(1-v^2_3)}\Bigr)^{-\frac{1}{2}}+
\int^1_{\sqrt{\frac{\nu}{\lambda}}} dv_3\, v^2_3\, \ep^{\lambda(1-v^2_3)}
\Bigl(1-v^4_3\,\ep^{2\lambda(1-v^2_3)}\Bigr)^{-\frac{1}{2}}
\biggr]
\,,
\end{equation}
with $\frac{\ep^{\lambda}}{\lambda}=\frac{\ep^{\nu}}{\nu}\cos\alpha_3$. If $\beta$ is a base angle of the isosceles triangle $Q_1Q_2O$, then a simple geometric analysis leads to 

\begin{equation}\label{Tt}
l=2\cos\beta\,l_1\,,
\qquad
L=\sin\beta\,l_1+l_3
\,.
\end{equation}

The energies of the two first strings can be read from \eqref{E+}, while that of the third from \eqref{E-}. Combining those with \eqref{baryon-v} for the gravitational energy of the vertex, we find the total energy of the configuration 

\begin{equation}\label{TE}
\begin{split}
E_{\3q}=&\g\sqrt{\frac{\s}{\nu}}
\biggl[
3\kappa\ep^{-2\nu}
+
2\int^1_0\,\frac{dv_1}{v_1^2}\,
\biggl(
\ep^{\nu v_1^2}\Bigl(1-\cos^2{}\hspace{-1mm}\alpha_1\,
v_1^4\,\ep^{2\nu (1-v_1^2)}\Bigr)^{-\frac{1}{2}}
-1-v_1^2
\biggr)
\biggr]\,+C\\
+&
\g\sqrt{\frac{\s}{\lambda}}
\biggl[
\int^1_0\,\frac{dv_3}{v_3^2}\,\biggl(\ep^{\lambda v_3^2}\Bigl(1-v_3^4\,\ep^{2\lambda (1-v_3^2)}\Bigr)^{-\frac{1}{2}}-1-v_3^2\biggr)
+
\int^1_{\sqrt{\frac{\nu}{\lambda}}}\,\frac{dv_3}{v_3^2}\,
\ep^{\lambda v_3^2}\Bigl(1-v_3^4\,\ep^{2\lambda (1-v_3^2)}\Bigr)^{-\frac{1}{2}}
\biggr]\,
\,.
\end{split}
\end{equation}

To complete the picture, we would need the gluing conditions at the vertex to show that the three-quark potential is written parametrically as $E_{\3q}=E(\nu,\lambda)$, $l=l(\nu,\lambda)$, and $L=L(\nu,\lambda)$. These conditions include two equations for force equilibrium, in the $y$ and $r$-directions. From \eqref{is-netforce} it follows that 

\begin{equation}\label{fbc}
\cos\alpha_3=2\sin\beta\cos\alpha_1\,,
\qquad
2\sin\alpha_1+\sin\alpha_3=3\kappa(1+4\nu)\ep^{-3\nu}
\,.
\end{equation}
Notice that $\alpha_3$ is negative.

Just as in the case of the collinear configuration with large $L$ and small $l$, we have to analyze the behavior of the parametric equations in a small region near the point $(0,1)$ in a two-dimensional parameter space $(\nu,\lambda)$. To this end, we choose the path $\nu=\varepsilon$ and $1-\sqrt{\lambda}=\varepsilon$ that enables us to have expansions in terms of $\varepsilon$ and control path-dependent terms. A straightforward but somewhat tedious calculation shows that 

\begin{equation}\label{limit}
\begin{split}
l=&\sqrt{\frac{\nu}{\s}}\Bigl(l_0+l_1\nu\Bigr)+o(\epsilon^{\frac{3}{2}})
\,,\quad
L=\frac{1}{\sqrt{\s}}\Bigl(-\ln(1-\sqrt{\lambda})+{\cal L}_0\Bigr)+o(\epsilon^{\oh})
\,,\\
E_{\3q}=&\g\sqrt{\s}\Bigl(-\ep\ln(1-\sqrt{\lambda})+{\cal E}_0\Bigr)+C
+\g\sqrt{\frac{\s}{\nu}}\Bigl(E_0+E_1\nu\Bigr)+o(\epsilon^{\oh})
\,.
\end{split}
\end{equation}
with the same $l_i$'s, ${\cal L}_0$, ${\cal E}_0$, and $E_i$'s as in \eqref{lLE-dq2}-\eqref{lLE-dq4}. As before, we indicate the dependence on $\nu$ and $\lambda$ rather than on $\varepsilon$ when no ambiguity arises. 

Finally, we reduce the parametric equations to a single one

\begin{equation}\label{Tdi-quark}
E_{\3q}(l,L)=-\frac{\alpha_{\qq}}{l}+\sigma_0l+\sigma L+\text{c}+o(l)\,,
\end{equation}
with the same $\alpha_{\qq}$, $\sigma_0$, $\sigma$, and $\text{c}$ as in \eqref{di-quark}.

As in Section III, we have to be careful in extracting the quark-quark potential from the above expression. In the limit we are considering, the $y$-coordinate of the vertex behaves like $y_{\scriptscriptstyle +}\sim\varepsilon^{\frac{3}{2}}\sim l^3$. This means that to the approximation we are using the projection of $V$ is located at the origin. In contrast, from the four-dimensional perspective it should be located at $y=l/2\sqrt{3}$ \cite{pdf-diquark}. Thus, we have to subtract from $E_{\3q}$ the term proportional to $\sigma$ and a constant term. As a result, we arrive at the expression \eqref{diquark}.

%___________________________________________________________________

\end{document}